%% file: main.tex
\documentclass[aps,pra,10pt,twocolumn,floatfix,superscriptaddress,longbibliography]{revtex4-2}

\usepackage{adjustbox}
\usepackage{nicematrix}
\usepackage{amssymb,amsmath,amstext,amsthm}
\usepackage{siunitx}
\usepackage{derivative}
\usepackage{csquotes}
\usepackage{array}
\usepackage{dsfont}
\usepackage{graphicx}
\usepackage{bm}
\usepackage{algorithm,algpseudocode}
\usepackage{appendix}
\usepackage[T1]{fontenc}
\usepackage{physics}
\usepackage{xcolor}
\usepackage[colorlinks=true,citecolor=blue,linkcolor=magenta]{hyperref}
\usepackage{tikz-network}
\usepackage{orcidlink}

\usepackage{tikz}
\usetikzlibrary{positioning, calc, decorations.pathreplacing}
\tikzset{tensor/.style={circle,draw,minimum size=10mm,inner sep=0.5mm}}

\newcommand{\re}{\mathrm{Re}}
\newcommand{\ma}{\mathrm{Ma}}

\AtBeginDocument{\RenewCommandCopy\qty\SI}

\newcommand{\mecheng}{Department of Mechanical Engineering, The University of Melbourne, Parkville, VIC 3010, Australia}
\newcommand{\schoolofphysics}{School of Physics, The University of Melbourne, Parkville, VIC 3010, Australia}
\newcommand{\quantummotionaus}{Quantum Motion, 4 Cornwallis St, Eveleigh NSW 2015, Australia}
\newcommand{\monash}{Faculty of Information Technology, Monash University Clayton 3800 Victoria Australia}

\begin{document}

\title{Quantum-Inspired Computational Fluid Dynamics for Transient Turbulent Compressible Flows}

\author{Shang Xian Matthew Lee\,\orcidlink{0009-0005-8307-7031}}
\affiliation{\mecheng}

\author{Melissa Kozul\,\orcidlink{0000-0001-9881-1677}}
\affiliation{\mecheng}

\author{Muhammad Usman\,\orcidlink{0000-0003-3476-2348}}
\affiliation{\monash}
\affiliation{\schoolofphysics}

\author{Martin Sevior\,\orcidlink{0000-0002-4824-101X}}
\affiliation{\schoolofphysics}

\author{Matthew L. Sims-Goh\,\orcidlink{0000-0002-7478-4026}}
\affiliation{\quantummotionaus}
\affiliation{\schoolofphysics}

\author{Richard D. Sandberg\,\orcidlink{0000-0001-5199-3944}}
\affiliation{\mecheng}

\begin{abstract}
Quantum-inspired algorithms are an emerging class of algorithms for computational fluid dynamics (CFD) with potentially favourable scaling for large problems compared to classical methods.
However, their applications have been limited to incompressible flows due to arithmetic limitations, which are addressed in this work.
This work introduces the first true end-to-end quantum-inspired computational fluid dynamics (QICFD) solver for direct numerical simulation of the compressible Navier--Stokes equations, that is, all arithmetic operations, as well as initialisation and post-processing, are undertaken in the tensor train (TT) format.
Importantly, new division and square-root algorithms using TTs enable the use of Sutherland's law for viscosity, and leverages recent improvements in TT arithmetic algorithms for reduced runtime and memory cost.
The new QICFD solver is validated by comparison with the classical CFD solver HiPSTAR and by way of a challenging fluid-flow test case, the low resolution Taylor--Green Vortex (TGV) at Mach numbers of \num{0.8} and \num{0.1}.
The TGV test case is a transient turbulent case that is very sensitive to accumulating errors, yet our QICFD solver achieves excellent agreement with the classical CFD reference.
This work demonstrates the correctness of the new TT division and square-root algorithms, and that QICFD is capable of compressible flow simulations.
The new QICFD solver is also able to perform simultaneous simulations, running multiple TGV-like cases initialised differently in parallel with reduced extra cost.
Finally, the demonstrated TGV test case reveals additional challenges of QICFD as well as highlight the need for future advances to make TT methods viable for industrially-relevant conditions.
\end{abstract}
\maketitle

\input{figures/summary/summary.tex}

\section{Introduction\label{sec:introduction}}

Computational fluid dynamics (CFD) is a crucial tool in the modern engineer's toolbox, allowing detailed evaluation of initial designs without the need for expensive and time consuming physical experiments.
Current industrial practice mainly relies on Reynolds Averaged Navier--Stokes (RANS) approaches for CFD, which are time-averaged analyses requiring less dense grids over ground-truth Direct Numerical Simulations (DNS)~\cite{sandberg2022fluid}.
This has significant implications for the accuracy and reliability of simulation results; for example RANS predictions for blade temperatures of a jet engine can give errors of up to \qty{400}{\kelvin} \cite{muller_EfficientUnsteadySimulations_2025}.
This implies that current designs must be very conservative to accommodate for simulation error, leaving room for further optimisation untapped.
There is a push towards high-fidelity CFD, such as Large Eddy Simulation (LES) or even DNS for more accurate simulations, however they are too computationally expensive for the parametric sweeps common in industrial design~\cite{sandberg2022fluid}. 
It is predicted that even computers of 2030 will be unable to perform LES of a single aircraft wing at true flight conditions, and it is unlikely algorithmic and classical hardware advancements will be able to meet the needs of current and future engineers \cite{slotnick_CFDVision2030_2014}.

The fundamental issue for CFD can be understood using the Reynolds number ($\re$), and how computational requirements scale with it.
$\re$ is the ratio of inertial forces versus viscous forces, and indicates the range of scales of turbulence in a fluid flow.
CFD simulations that wish to accurately represent turbulence must resolve all relevant time and length scales, which inform the grid size and time step requirements.
\textcite{yang_GridpointTimestepRequirements_2021} investigated the scaling behaviour of DNS and LES for the Navier--Stokes equations, reporting a power law scaling of the grid point requirements as $\re^{2.05}$ for DNS and $\re^{1.86}$ for wall-resolved LES and a time-step scaling of $\re^{0.857}$ for both DNS and wall-resolved LES.
$\re$ can range from about one thousand for insect wings to about one billion for atmospheric flows, and the strong scaling of computational requirements make high-fidelity CFD of large and complicated problems prohibitively expensive.
For example, DNS of a single turbine blade within a jet engine of a commercial jet airliner already requires hundreds of GPUs and over 1 week of wall time \cite{nardini_DirectNumericalSimulation_2024}, noting a complete engine contains thousands of blades, and even that is only a small portion of the entire aircraft.
The high computational cost for complex flows means it is unlikely traditional methods will be capable of performing high-fidelity CFD for demanding industrial use cases in the near future.
To overcome this computational limitation, new innovative methods that scale better are needed.

An emerging approach is quantum-inspired CFD (QICFD), which borrows tools from the simulation of quantum systems as the basis for an unconventional approach to CFD.
Tensor network methods are an enormously successful tool for quantum simulation, and have played a crucial role in quantum science for decades~\cite{white1992density,vidal2003efficient,vidal2004efficient,perez2006matrix,verstraete2008matrix,schollwock2011density,orus2014practical,cirac2021matrix,nakhl_StabilizerTensorNetworks_2025,sims2026multivariate}.
These methods represent exponentially-sized multipartite quantum states as a product network of smaller local tensors, which can be manipulated directly without ever constructing the full state.
When the entanglement structure of the state is limited, tensor network methods can represent the state with exponentially fewer parameters, and perform certain operations on the state with exponentially less computational cost~\cite{verstraete2006matrix,schuch2008entropy,eisert2010colloquium}.
The simplest and most popular tensor network is the tensor train (TT)~\cite{oseledets2011tensor,grasedyck2013literature}, commonly known as a matrix product state in the quantum simulation community \cite{perez2006matrix,schollwock2011density,orus2014practical,cirac2021matrix}; we restrict our attention to this type of tensor network in this work.

In recent years there has also been an interest in tensor network methods applied to problems completely unrelated to quantum mechanics, such as the compressed representation and manipulation of multivariate functions.
Discretised multivariate functions unrelated to quantum mechanics can be encoded as tensor networks; just as weakly-entangled quantum states can admit an efficient representation in this form~\cite{verstraete2006matrix,schuch2008entropy,eisert2010colloquium}, sufficiently well-behaved functions are also exponentially compressible in this form~\cite{oseledets2010approximation,khoromskij_OdlogNQuanticsApproximation_2011}.
By manipulating the tensors directly in this encoding, one can in principle directly simulate a variable's evolution under a partial differential equation --- this has previously been applied to a range of nonlinear dynamics problems including incompressible CFD~\cite{gourianov_QuantumInspiredApproach_2022, pisoni_CompressionSimulationSynthesis_2026}, nonlinear plasma dynamics~\cite{ye2022quantum}, and Bose--Einstein condensate dynamics~\cite{connor2026tensor,bou2025quantics,niedermeier2026solving,chen2025solving}.
In CFD, TT methods were first used for QICFD by \textcite{gourianov_QuantumInspiredApproach_2022}, who applied TT methods to solve the incompressible Navier--Stokes equations, simulating the 2D temporally developing jet and the 3D Taylor--Green Vortex test cases.
QICFD outperformed classical CFD in terms of accuracy for the same level of memory use, which points to the potential computational advantage of QICFD.
Subsequent works aimed to improve the performance of QICFD through algorithmic improvements and GPU acceleration \cite{holscher_QuantuminspiredFluidSimulation_2025, pham_RandomizedTensorTrainTruncation_2026}, and introduced more features to QICFD, such as solid boundaries \cite{kiffner_TensorNetworkReduced_2023,peddinti_QuantuminspiredFrameworkComputational_2024}, finite element analysis \cite{kornev_TetraFEMNumericalSolution_2024}, curvilinear coordinates \cite{hulst_QuantumInspiredTensorNetworkFractionalStep_2025}, buoyancy driven flows \cite{hulst_QuantumInspiredSimulation2D_2026}, and also alternate formulations such as using the lattice Boltzmann method \cite{gross_TensorNetworkLattice_2026}.

Despite significant research interest in nonlinear classical simulations with TT methods, existing work on QICFD has been mostly limited to incompressible flows.
Compressible flows are those when the changes in fluid density are significant: for example, in aerospace applications when the flow velocity approaches the speed of sound~\cite{nardini_DirectNumericalSimulation_2024}, and when studying engineering acoustics, one must explicitly simulate pressure waves \cite{wang_EffectCompressibilityFlowinduced_2025}.
However, simulating compressible flow presents a challenge for existing QICFD methods.
To avoid constructing the intractable (exponentially-large) full state vector, one must perform all required operations directly on the individual tensors of the TT.
Not all operations can be generalised to this form, and previous work has provided routines for basic arithmetic such as addition, multiplication, and applying linear operators \cite{lee_FundamentalTensorOperations_2016}, and more advanced operations such as Fourier transforms, finite differences and integration \cite{garcia-ripoll_QuantuminspiredAlgorithmsMultivariate_2021}.
However, the compressible version of the Navier--Stokes equations requires element-wise division (since the density is no longer constant) and square-root operations (to calculate the now inhomogeneous viscosity using Sutherland's law \cite{white_ViscousFluidFlow_1991}).
Prior work has been greatly limited by this --- the limited existing investigations have either explicitly contracted the exponentially-sized full state vector \cite{danis_TensortrainWENOScheme_2025} (sacrificing the end-to-end scaling advantage of QICFD) or been bottlenecked by highly inefficient division algorithms that are much more costly than other dominant elementary operations \cite{pinkston_MatrixProductState_2025}.
Preceding works also made the assumption of constant viscosity, which is unrealistic for compressible flows.

The present effort overcomes these limitations by introducing accurate and efficient division and square-root algorithms in the TT format.
This enables us to develop a complete compressible QICFD solver that is highly efficient and scalable to large problems.
We validate our compressible QICFD solver by comparison to a state-of-the-art classical solver HiPSTAR \cite{sandberg_CompressibleDirectNumerical_2015}.
The QICFD solver achieves excellent agreement with the classical reference, showing the validity of the new division and square-root algorithms, and the QICFD solver overall.
We also demonstrate a novel method of conducting simultaneous simulations in the TT format, which shows the same scaling behaviour with increasing problem size.
This provides another pathway for QICFD to out-scale classical CFD, by combining classically-parallel simulations into a simultaneous simulation with only a reduced cost increase over a single simulation, enabling an efficient and scalable method of running massively-parallel simulations.
A summary of our work, and the overall end-to-end workflow of our QICFD solver in presented in Figure \ref{fig:summary}.
Our contributions represent a significant advancement in the capabilities of quantum-inspired simulation in the field of compressible CFD.

\section{Methods} \label{sec:methods}

\subsection{Tensor Train Basics}
In this work, we use tensor networks to encode the terms of a compressible CFD simulation using a quantum-inspired method running on classical hardware.
In the quantum simulation context, an $N$-qubit quantum state is described by a state vector $\ket{\psi} \in \mathcal{H}^{\otimes N}$, where $\mathcal{H} \cong \mathbb{C}^2$ is the Hilbert space of a single qubit.
We use standard amplitude encoding~\cite{zalka1998simulating,grover2002creating} in the \enquote{quantics} representation~\cite{khoromskij_OdlogNQuanticsApproximation_2011,shinaoka2023multiscale,ritter2024quantics,fernandez_LearningTensorNetworks_2025} --- for a univariate function $f:[0,1) \to \mathbb{R}$, the function is discretised on a uniform grid of size $2^N$ at grid points $x_j = j/2^N$ for $j=0,\dots,2^N-1$, and encoded as the state
\begin{equation}
  \ket{\psi}=\sum_j f(x_j) \ket{j}, \quad x_j \equiv \sum_{b=1}^{N}\frac{q_{b}}{2^b} = \underbrace{0.q_{1} q_{2} \dots q_{N}}_{\text{binary representation}},\label{eqn:quantics_representation}
\end{equation}
where $\ket{j} \equiv \ket{q_{1} q_{2} \dots q_{N}}$ indexes the grid points in binary counting order.
A general bounded domain can be accommodated by a simple linear rescaling.
Multivariate functions can be accommodated via the product of several such registers: a $D$-dimensional function $f(x_1,\dots,x_D)$ discretised on a $2^P$-point grid in each dimension, is encoded using one $P$-qubit register per variable as
\begin{equation}
\ket{\psi} = \sum_{j_1,\dots,j_D} f(x_{j_1},\dots,x_{j_D}) \ket{j_1}\otimes\cdots\otimes\ket{j_D} \in \mathcal{H}^{\otimes PD}.\label{eqn:multivariate_psi}
\end{equation}
Naively this encoding scales intractably, as the state $\ket{\psi}$ requires $2^{PD}$ complex amplitudes to describe --- however if it is written as the product of smaller tensors contracted together, a suitably structured $f$ may be represented with exponentially fewer parameters~\cite{verstraete2006matrix,schuch2008entropy,eisert2010colloquium,oseledets2010approximation,khoromskij_OdlogNQuanticsApproximation_2011}.
Such a product is known as a tensor network; the simplest and most popular choice to represent a state $\ket{\psi}$ is the TT (also known as Matrix Product States MPS)~\cite{oseledets2011tensor,grasedyck2013literature,perez2006matrix,schollwock2011density,orus2014practical,cirac2021matrix}.
Decomposing the amplitudes of $\ket{\psi}$ into a train of tensors $A^1,\dots,A^{N}$ with $N=PD$, the TT encoding of Equation~\ref{eqn:multivariate_psi} is given by
\begin{align}
  f_{j_1,\dots,j_D} &= \prod_{n=1}^{PD} [A^n]^{q_n} \\
  &= [A^{(1,1)}]^{q_{1,1}}_{\alpha_1} [A^{(1,2)}]^{q_{1,2}}_{\alpha_1\alpha_2} \cdots  [A^{(D,P)}]^{q_{D,P}}_{\alpha_{PD-1}}, \label{eqn:tt_encoding}
\end{align}
where the Einstein summation convention is assumed, and the tensors are indexed by position $n = (k-1)D+b$, and $j_n \equiv j_{k,b}$ indexes the Hilbert space $\mathcal{H}$ of the corresponding qubit.
Unlike a truly quantum algorithm, the states are stored entirely in classical memory, so there is no need to enforce $\ell^2$ normalisation.

The indices $\alpha_n$ represent the \enquote{virtual indices} of the TT, which enable correlation of different parts of the function $f$ across different variable values (i.e. entanglement in the quantum state $\ket{\psi}$).
As a consequence of the Schmidt decomposition at each virtual index, $\operatorname{dim}(\alpha_n)\leq2^{\min(n, N-n)}$ --- that is, its dimension can in general grow exponentially in the distance from the nearest end of the chain.
When this virtual index is saturated, the TT representation offers no advantage over the standard dense representation in Equation~\ref{eqn:multivariate_psi}.
Singular value decomposition (SVD) is a common method to encode functions into the TT format, in which case $\alpha_n$ directly corresponds to the number of singular values (denoted $\sigma$) retained.
During encoding, small singular values (typically singular values smaller than a fraction of the largest singular value $\sigma < \sigma_c \operatorname{max}(\sigma)$) can be truncated, which is equivalent to finding an approximation of the encoded data.
Commonly, one quantifies the complexity of the TT by the largest such dimension $\chi = \max_n(\operatorname{dim}(\alpha_n))$, known as the \enquote{rank} in TT literature or the \enquote{bond dimension} in MPS literature.
If $\chi$ is the typical size of $\alpha_n$, then approximately only $2N\chi^2$ parameters are needed to encode the original data of size $2^{N}$, representing an exponential memory saving.
As noted in Section~\ref{sec:introduction}, $\chi$ of the TT may be exponentially smaller for weakly-entangled quantum states~\cite{verstraete2006matrix,schuch2008entropy,eisert2010colloquium} or well-behaved functions at controllable error tolerance~\cite{oseledets2010approximation,khoromskij_OdlogNQuanticsApproximation_2011}.
Therefore, manipulating $f$ in the TT format rather than enumerating all $2^{N}$ amplitudes can expose an exponentially more efficient representation, enabling enormous asymptotic speedups for computations on $f$ that can be efficiently performed on the individual tensors $A_n$.
For instance, one can perform linear algebra operations such as element-wise addition, operator-vector products \cite{lee_FundamentalTensorOperations_2016}, and ultra-fast Fourier transforms~\cite{chen2023quantum} in this encoded form.
In general, the TT algorithm used in this work has runtime and memory scaling of $O(N\chi^4)$ and $O(N\chi^3)$.
Provided $\chi$ is small for the application, the TT format can provide a scaling advantage over operating on the original dense-field data.

Tensor networks are often most easily understood in Penrose graphical notation~\cite{penrose1971applications,bridgeman2017hand}: tensors are represented as nodes connected by lines, internal lines connecting two nodes correspond to summed-over \enquote{virtual} indices, whilst external lines with a free end correspond to local Hilbert space \enquote{physical} indices.
In this notation, Equation \ref{eqn:tt_encoding} can be re-written as
\begin{equation}
f_{j_1,j_2,j_3} =
\begin{array}{c}
\begin{tikzpicture}
  \Vertex[x=0,   y=0,label=$A^{(1,1)}$,position=above,size=0.25,color=blue!30]{A11}
  \Vertex[x=1.2, y=0,label=$A^{(1,P)}$,position=above,size=0.25,color=blue!30]{A1N}
  \Vertex[x=2.1,y=0,label=$A^{(2,1)}$,position=above,size=0.25,color=orange!60]{A21}
  \Vertex[x=3.3,y=0,label=$A^{(2,P)}$,position=above,size=0.25,color=orange!60]{A2N}
  \Vertex[x=4.2, y=0,label=$A^{(3,1)}$,position=above,size=0.25,color=green!60]{A31}
  \Vertex[x=5.4, y=0,label=$A^{(3,P)}$,position=above,size=0.25,color=green!60]{A3N}

  \draw[line width=1] (A11) -- ++(0,-0.35) node[below] {\scriptsize$q_{1,1}$};
  \draw[line width=1] (A1N) -- ++(0,-0.35) node[below] {\scriptsize$q_{1,P}$};
  \draw[line width=1] (A21) -- ++(0,-0.35) node[below] {\scriptsize$q_{2,1}$};
  \draw[line width=1] (A2N) -- ++(0,-0.35) node[below] {\scriptsize$q_{2,P}$};
  \draw[line width=1] (A31) -- ++(0,-0.35) node[below] {\scriptsize$q_{3,1}$};
  \draw[line width=1] (A3N) -- ++(0,-0.35) node[below] {\scriptsize$q_{3,P}$};

  \Edge[style={dashed},label={\makebox[1em][c]{$\cdots$}}](A11)(A1N)
  \Edge[style={dashed},label={\makebox[1em][c]{$\cdots$}}](A21)(A2N)
  \Edge[style={dashed},label={\makebox[1em][c]{$\cdots$}}](A31)(A3N)

  \Edge[lw=1,position=below,label={\makebox[1em][c]{$\alpha_P$}}](A1N)(A21)
  \Edge[lw=1,position=below,label={\makebox[1em][c]{$\alpha_{2P}$}}](A2N)(A31)
\end{tikzpicture}
\end{array},
\label{eqn:tt_diagram}
\end{equation}
where the dashed lines represent the $P-2$ notationally suppressed tensors in each register, and we have additionally assumed $D=3$ since we target 3D CFD problems in this work. 
Noting that each of the three colours in Equation~\ref{eqn:tt_diagram} corresponds to a given spatial dimension, the graphical representation exposes an implicit assumption on the ordering of indices in the TT structure.
For univariate functions one simply places the physical indices in order of bit significance, but for multivariate functions one generally choose between \enquote{interleaved} (e.g. $x_1,y_1,z_1,x_2,y_2,z_2,\dots$) or \enquote{serial} (e.g. $x_1,\dots,x_P,y_1,\dots,y_P,z_1,\dots,z_P$) orderings of the physical indices \cite{ye2022quantum,gourianov_QuantumInspiredApproach_2022,rodriguez2024chebyshev,pisoni_CompressionSimulationSynthesis_2026}.
In this work, we exclusively use serial ordering since it allows operators that operate on distinct spatial dimensions (such as derivative operators) to be easily constructed and applied.

Linear operators acting on $\mathcal{H}^{\otimes PD}$ admit an analogous factorisation. Writing an operator $O$ in the computational basis of Equation~\ref{eqn:multivariate_psi} as $O = \sum_{\bm{j},\bm{j}'} O^{\bm{j}'}_{\bm{j}}\ket{j'_1,\dots,j'_D}\bra{j_1,\dots,j_D}$ (where $\bm{j}=(j_1,\dots,j_D)$, the matrix elements $O_{\bm{j}}^{\bm{j}'}$ form a $2^{PD}\times 2^{PD}$ tensor that can be factorised as
\begin{equation}
O^{\bm{j}'}_{\bm{j}} =
\begin{array}{c}
\begin{tikzpicture}
  \Vertex[x=0,   y=0,size=0.25,color=blue!30]{W11}
  \Vertex[x=1.2, y=0,size=0.25,color=blue!30]{W1N}
  \Vertex[x=2.1, y=0,size=0.25,color=orange!60]{W21}
  \Vertex[x=3.3, y=0,size=0.25,color=orange!60]{W2N}
  \Vertex[x=4.2, y=0,size=0.25,color=green!60]{W31}
  \Vertex[x=5.4, y=0,size=0.25,color=green!60]{W3N}

  \draw[line width=1] (W11) -- ++(0,0.35) node[midway,right] {\scriptsize$q'_{1,1}$};
  \draw[line width=1] (W1N) -- ++(0,0.35) node[midway,right] {\scriptsize$q'_{1,P}$};
  \draw[line width=1] (W21) -- ++(0,0.35) node[midway,right] {\scriptsize$q'_{2,1}$};
  \draw[line width=1] (W2N) -- ++(0,0.35) node[midway,right] {\scriptsize$q'_{2,P}$};
  \draw[line width=1] (W31) -- ++(0,0.35) node[midway,right] {\scriptsize$q'_{3,1}$};
  \draw[line width=1] (W3N) -- ++(0,0.35) node[midway,right] {\scriptsize$q'_{3,P}$};

  \draw[line width=1] (W11) -- ++(0,-0.35) node[midway,right] {\scriptsize$q_{1,1}$};
  \draw[line width=1] (W1N) -- ++(0,-0.35) node[midway,right] {\scriptsize$q_{1,P}$};
  \draw[line width=1] (W21) -- ++(0,-0.35) node[midway,right] {\scriptsize$q_{2,1}$};
  \draw[line width=1] (W2N) -- ++(0,-0.35) node[midway,right] {\scriptsize$q_{2,P}$};
  \draw[line width=1] (W31) -- ++(0,-0.35) node[midway,right] {\scriptsize$q_{3,1}$};
  \draw[line width=1] (W3N) -- ++(0,-0.35) node[midway,right] {\scriptsize$q_{3,P}$};

  \Edge[style={dashed},label={\makebox[1em][c]{$\cdots$}}](W11)(W1N)
  \Edge[style={dashed},label={\makebox[1em][c]{$\cdots$}}](W21)(W2N)
  \Edge[style={dashed},label={\makebox[1em][c]{$\cdots$}}](W31)(W3N)

  \draw[line width=1] (W1N) -- (W21) node[midway,below=0.22cm]{};
  \draw[line width=1] (W2N) -- (W31) node[midway,below=0.22cm]{};
\end{tikzpicture}
\end{array}.
\label{eqn:tto_diagram}
\end{equation}
In the case of special structure such as low-entanglement quantum operations~\cite{osborne2006efficient,prosen2007operator,pirvu2010matrix} or scale-separation-preserving operations like finite differences \cite{garcia-ripoll_QuantuminspiredAlgorithmsMultivariate_2021} or the Fourier transform~\cite{chen2023quantum}, this factorisation may be of low rank.
One can apply a TTO to a TTS using contractions, which is equivalent to matrix-vector multiplication, and various algorithms exists for this fundamental operation \cite{lee_FundamentalTensorOperations_2016, paeckel_TimeevolutionMethodsMatrixproduct_2019, michailidis_ElementwiseMultiplicationTensor_2025, meng_RecursiveSketchedInterpolation_2026, camano_SuccessiveRandomizedCompression_2026, milbradt_EfficientApplicationTensor_2026}.
In this form, the advantage of the serial representation is manifestly visible: one can easily apply single-variable TTOs %
\begin{tikzpicture}[baseline=-0.55ex]
  \Vertex[x=0,    y=0,size=0.16,color=blue!30,style={line width=0.5pt}]{ta}
  \Vertex[x=0.45, y=0,size=0.16,color=blue!30,style={line width=0.5pt}]{tb}
  \Vertex[x=1.3, y=0,size=0.16,color=blue!30,style={line width=0.5pt}]{tc}
  \foreach \n in {ta,tb,tc}{%
    \draw[line width=0.55pt] (\n) -- ++(0.15,0.15);
    \draw[line width=0.55pt] (\n) -- ++(-0.15,-0.15);}
  \Edge[lw=0.55](ta)(tb)
  \Edge[lw=0.55,style={dotted}](tb)(tc)
\end{tikzpicture}, such as the finite difference \cite{garcia-ripoll_QuantuminspiredAlgorithmsMultivariate_2021}, by contracting them directly at their relevant section of the TT.

To ease the discussion below, the symbols and nomenclature used in this work are collected in short in Table \ref{tab:short symbols} and in full in Table \ref{tab:full symbols}.

\begin{table*}[htb]
    \centering
    \begin{NiceTabular}{l|p{5cm}|p{10cm}}
        Symbol & Full name & Additional comments \\
        \hline \hline
         & Dense form & Original classical form of data \\
        TT & Tensor Train & Refers to both TTS and TTO \\
        TTS & Tensor Train State & TT form of a vector (data) (Matrix Product State (MPS) in MPS context) \\
        TTO & Tensor Train Operator & TT form of a matrix (linear operator) (Matrix Product Operator (MPO) in MPS context) \\
        $N$ & Total number of TT tensors & \\
        $q_{n}$ & Physical index & Physical index $n$ (or dimension of) (site in MPS context) \\
        $\alpha_n$ & Virtual index & Virtual index $n$ (or dimension of) (bond in MPS context) \\
        $\sigma_c$ & Relative singular value cutoff & Cutoff relative to the largest singular value during Singular Value Decomposition (SVD) truncation \\
        $\chi$ & Rank & Maximum dimension of virtual indices (bond dimension in MPS context) \\
        $\odot$ & Element-wise multiply & \\
        $\re$ & Reynolds number & Ratio of inertial forces over viscous forces \\
        $\ma$ & Mach number & Characteristic flow velocity over speed of sound \\
    \end{NiceTabular}
    \caption{Important symbols and nomenclature. The full table of symbols is in Table \protect{\ref{tab:full symbols}}.}
    \label{tab:short symbols}
\end{table*}

\subsection{Tensor Train Arithmetic Algorithms} \label{sec:arithmetics}

The main innovations in this work are robust and accurate division and square-root algorithms.
The two algorithms are conceptually very simple, and use the Newton--Raphson method.
During the final preparation of this manuscript, a pre-print describing the same algorithms was released \cite{wang_IterativeTensorNetwork_2026}.
We strongly encourage the interested reader to consult both works for completeness.

First is the division algorithm.
Division is split into reciprocation of the divisor, then multiplication of the result with the dividend.
The Newton--Raphson method is used to reciprocate the divisor.
The equation to solve and the corresponding iterations are
\begin{gather}
  f(x) = \frac{1}{x} - D = 0, \\
  x_{i+1} = 2x_i - Dx_i^2,
\end{gather}
where $x$ is the desired root, and $D$ is the divisor.
Now reciprocation, and hence division is reduced to scalar multiplication, element-wise addition, and element-wise multiplication, which are well known in the TT format.

The Newton--Raphson method requires an initial starting point, and a second order optimal polynomial approximation for the interval $[a, b]$ is used \cite{walczyk_OptimalApproximation1_2025}
\begin{equation}
  \frac{1}{x} \approx \frac{6 \left( 3a^{2}+ 10ab + 3b^{2} \right) - 48 \left( a + b \right) x + 32x^{2}}{\left( a + b \right) \left( a^{2} + 14ab + b^{2} \right)}
\end{equation}
It is trivial to guess an appropriate interval for CFD applications, especially when non-dimensional values are used.
A terminating condition is also needed, and the suggested metric to use is
\begin{equation}
  \varepsilon = \left| \mathbf{x}_i \cdot \mathbf{D} - 2^N \right| / 2^N,
\end{equation}
where $\mathbf{x}_i \cdot \mathbf{D}$ is the inner product of the two TTSs, and $2^N$ is the number of elements encoded.

There exists a previous division algorithm used by \textcite{pinkston_MatrixProductState_2025} termed the \enquote{DMRG}-like algorithm \cite{oseledets_SolutionLinearSystems_2012}.
This method has runtime and memory scaling of $O(N\chi^4)$, and can be considered a local optimisation method, iteratively fitting individual tensors to approach the optimal solution.
The main disadvantage of this method is its poor performance and unreliability.
Being an optimisation method, it may get stuck in local optimums and fail to produce a correct result, and it may also require many iterations.
The new algorithm is more robust and performant than the previous method, and benefits from recent improvements of multiplication algorithms which achieve $O(N\chi^3)$ memory scaling.
This is a significant improvement and an enabler over the previous method, as $\chi$ used in this work can reach 128 or 256 which already proves problematic for current generation hardware.

\begin{figure}
\begin{algorithm}[H]
\caption{Element-wise TT division} \label{alg:division}
\begin{algorithmic}[1]
    \Require $0 < a < b$
    \Require $a \leq \operatorname{minumum}(\mathbf{D})$
    \Require $b \geq \operatorname{maximum}(\mathbf{D})$
    \Procedure{DivideTT}{$\mathbf{N}$, $\mathbf{D}$, $a$, $b$, $\varepsilon_{tolerance}$, $i_{max}$}
        \State $p_0 \gets \frac{3a^{2}+ 10ab + 3b^{2}}{\left( a + b \right) \left( a^{2} + 14ab + b^{2} \right)}$
        \State $p_1 \gets \frac{- 48 \left( a + b \right)}{\left( a + b \right) \left( a^{2} + 14ab + b^{2} \right)}$
        \State $p_2 \gets \frac{32}{\left( a + b \right) \left( a^{2} + 14ab + b^{2} \right)}$
        \State $\mathbf{x} \gets p_0 + p_1 \mathbf{D} + p_2 \mathbf{D} \odot \mathbf{D}$
        \State $i \gets 1$
        \State $\varepsilon \gets \Call{CalculateError}{\mathbf{x}, \mathbf{D}}$
        \While{$i < i_{max}$ and $\varepsilon > \varepsilon_{tolerance}$}
            \State $\mathbf{x} \gets 2\mathbf{x} - \mathbf{D} \odot \mathbf{x} \odot \mathbf{x}$
            \State $\varepsilon \gets \Call{CalculateError}{\mathbf{x}, \mathbf{D}}$
            \State $i \gets i + 1$
        \EndWhile
        \State $1/\mathbf{D} \gets \mathbf{x}$
        \State \Return $1/\mathbf{D} \odot \mathbf{N}$
    \EndProcedure
\end{algorithmic}
\end{algorithm}
\end{figure}

A novel square-root algorithm is also specifically developed in this work.
This is the first algorithm for square-roots in the TT format.
The new square-root algorithm also uses the Newton--Raphson method.
The equation to solve and the corresponding iterations are
\begin{gather}
  f(x) = x^2 - D = 0, \\
  x_{i+1} = \frac{1}{2} \left( x_i + \frac{D}{x_i} \right),
\end{gather}
where $x$ is the desired root, and $D$ is the input.
This algorithm now requires scalar multiplication and addition which are well known, and division which can be calculated as introduced above.
Any initial condition will work, and generally using the input $\mathbf{D}$ is sufficient.
The suggested metric for a termination condition is
\begin{equation}
  \varepsilon = \left| \lvert \lvert \mathbf{x}_i \rvert \rvert^2 - \mathrm{sum}(\mathbf{D}) \right| / 2^N,
\end{equation}
where $\lvert \lvert \mathbf{x} \rvert \rvert$ is the norm of the TTS, and $\mathrm{sum}(\mathbf{D})$ is the sum of all elements of the TTS, which can be calculated using the same method as integration of a TTS \cite{garcia-ripoll_QuantuminspiredAlgorithmsMultivariate_2021}.

\begin{figure}
\begin{algorithm}[H]
\caption{Element-wise TT square-root} \label{alg:square root}
\begin{algorithmic}[1]
    \Procedure{SqrtTT}{$\mathbf{D}$, $\varepsilon_{tolerance, sqrt}$, $i_{max, sqrt}$, $a$, $b$, $\varepsilon_{tolerance, div}$, $i_{max, div}$}
        \State $\mathbf{x} \gets \mathbf{D}$
        \State $i \gets 1$
        \State $\varepsilon \gets \Call{CalculateError}{\mathbf{x}, \mathbf{D}}$
        \While{$i < i_{max, sqrt}$ and $\varepsilon > \varepsilon_{tolerance, sqrt}$}
            \State $\mathbf{D}/\mathbf{x} \gets \Call{DivideTT}{\mathbf{D}, \mathbf{x}, a, b, \varepsilon_{tolerance, div}, i_{max, div}}$
            \State $\mathbf{x} \gets 0.5 \left( \mathbf{x} + \mathbf{D}/\mathbf{x} \right)$
            \State $\varepsilon \gets \Call{CalculateError}{\mathbf{x}, \mathbf{D}}$
            \State $i \gets i + 1$
        \EndWhile
        \State \Return $\mathbf{x}$
    \EndProcedure
\end{algorithmic}
\end{algorithm}
\end{figure}

\begin{table*}
  \centering
  \begin{NiceTabular}{p{4cm}|p{4cm}|p{3cm}|p{3cm}}
    Operation & Algorithm & Runtime scaling & Memory Scaling \\
    \hline \hline
    Scalar multiplication & Trivial & $O(1)$ & None \\
    Addition & Density matrix \cite{fishman_ITensorSoftwareLibrary_2022} & $O(N\chi^3)$ & $O(\chi^2)$ \\
    Addition & DMRG-like \cite{paeckel_TimeevolutionMethodsMatrixproduct_2019} & $O(N\chi^3)$ & $O(N\chi^2)$ \\
    TTS-TTO contraction & CBC \cite{milbradt_EfficientApplicationTensor_2026} & $O(N\chi^4)$ & $O(N\chi^3)$ \\
  \end{NiceTabular}
  \caption{List of TT algorithms used in this work} \label{tab:TT algo}
\end{table*}
Other TT operations used are listed in Table \ref{tab:TT algo}.
All other algebraic operations, such as multiplication, division, square-roots, and finite differences (see Appendix \ref{sec:FDTTO}) can be expressed as a combination of the above fundamental operations.
In particular, element-wise multiplication is performed by first converting a TTS into a TTO using \enquote{delta} tensors \cite{hulst_QuantumInspiredTensorNetworkFractionalStep_2025}, then TTS-TTO contraction is performed.
To the authors' knowledge, this work is the first application of the Cholesky-Based Compression (CBC) algorithm \cite{milbradt_EfficientApplicationTensor_2026}, and it is found to be superior to previous contraction algorithms in wall-time and memory use.
Two addition algorithms are used, the density matrix addition algorithm and the DMRG-like addition algorithm.
The density matrix addition algorithm is used for most additions, given its critical ability to sum multiple terms in a single step.
The DMRG-like addition algorithm is reserved for the addition within the division algorithm, as the density matrix addition algorithm was found to be inaccurate for this particular addition, an issue also encountered by \textcite{camano_SuccessiveRandomizedCompression_2026} (for TTS-TTO contraction).
The final scaling of the entire application is determined by the worst-scaling arithmetic algorithm, which is $O(N\chi^4)$ runtime and $O(N\chi^3)$ memory for TTS-TTO contraction.

\subsection{Quantum-Inspired CFD Solver} \label{sec:QICFD}

With the new division and square-root algorithms, all algebraic operations required for solving the compressible Navier--Stokes equations are available, enabling the development of a complete QICFD solver in this work.
This work demonstrates the first full QI 3D compressible CFD solver.
The solver closely mimics HiPSTAR \cite{sandberg_CompressibleDirectNumerical_2015}, an in-house state-of-the-art classical finite-difference CFD solver developed within the authors' group.
The QICFD solver is described in detail in Appendix \ref{sec:solver details}.

The Navier--Stokes equations for a compressible flow in conservative derivative form with skew-symmetric splitting of derivatives \cite{kennedy_ReducedAliasingFormulations_2008} are used:
\begin{align}
    \pdv{\rho}{t} &= - \nabla_j \cdot \left( \alpha_q \rho \mathbf{u}_j \right)\
    - \beta_q \left( \rho \nabla_j \cdot \mathbf{u}_j + \mathbf{u}_j \cdot \nabla_j \rho \right) \\
    &\begin{aligned}
        \pdv{\rho \mathbf{u}_i}{t} &= - \nabla_j \cdot \left( \alpha \rho \mathbf{u}_i \mathbf{u}_j + p \delta_{ij} - \mathbf{\tau}_{ji} \right) \\
        & \qquad - \gamma \left( \mathbf{u}_i \mathbf{u}_j \cdot \nabla_j \rho + \rho \mathbf{u}_j \cdot \nabla_j \mathbf{u}_i \right. \\
        & \qquad \qquad \left. + \rho \mathbf{u}_i \nabla_j \cdot \mathbf{u}_j \right)
    \end{aligned} \\
    &\begin{aligned}
        \pdv{\rho e}{t} &= - \nabla_j \cdot \left( \alpha \rho e \mathbf{u}_j + \alpha_q p \mathbf{u}_{j} - \mathbf{\tau}_{ji} \cdot \mathbf{u}_{i} + \mathbf{q}_j \right) \\
        & \qquad - \gamma \left( e \mathbf{u}_j \cdot \nabla_j \rho + \rho \mathbf{u}_j \cdot \nabla_j e + \rho e \nabla_j \cdot \mathbf{u}_j \right) \\
        & \qquad - \beta_q \left( p \nabla_j \cdot \mathbf{u}_j + \mathbf{u}_j \cdot \nabla_j p \right)
    \end{aligned}
\end{align}
$\rho$ is the fluid density, $\mathbf{u}$ is the fluid velocity vector, and $e$ the fluid specific energy, $\mathbf{\tau}$ is the viscous stress tensor, $\mathbf{q}$ is the heat flux vector, and $p$ is the fluid pressure, $\nabla$ is the derivative operator, $\delta$ is the Kronecker delta, $i,j$ are spatial direction indices, $\alpha_q, \beta_q, \alpha, \gamma$ are coefficients \cite{kennedy_ReducedAliasingFormulations_2008}.
An expanded version of the equations and details of each term are presented in Appendix \ref{sec:ns eq}.

Both solvers perform direct numerical simulations (DNS), i.e. the full Navier--Stokes equations are solved without any modelling, hence requiring resolution of all spatial and temporal scales, which is the most accurate type of CFD simulation.
The RK4 time stepping scheme is used, which uses a weighted average of four time derivatives using the right-hand side of the equations per physical time step for fourth-order time accuracy.
Finite difference solvers discretise the problem on a structured grid, in this work a uniform 3D Cartesian grid, and quantities are calculated at the nodes of the grid.
Spatial derivatives are discretised using fourth-order central difference stencils, which read as:
\begin{equation}
    \left. \pdv{u}{x} \right|_{x = i} \approx \frac{-u_{i+2} + 8 u_{i+1} - 8u_{i-1} + u_{i-2}}{12 \Delta x}
\end{equation}
where $u$ is the quantity for which the derivative is sought and $i$ is the index of the grid cell.
Most of the features of the solver, such as RK4 time stepping, fourth-order spatial derivatives, and skew-symmetric splitting of derivatives improve the stability and accuracy of the solver, as small errors will accumulate quickly in a transient simulation.
These features come at a high computational cost, for instance RK4 time stepping calculates time derivative using the Navier--Stokes equations four times per physical time step.
Skew-symmetric splitting of derivatives introduces forty eight new terms to the equations, however the method significantly improves the numerical stability of simulating the full non-linear equations for compressible cases.

The following briefly demonstrates how division and square-roots are used in our QICFD solver.
Division is used for updating the primitive variables, such as $u$ (velocity in $x$-direction).
Our CFD solver uses the conservative formulation of the Navier--Stokes equations which produces derivatives such as $\pdif*{\rho u} / \pdif*{t}$, which updates $\rho u$, which then needs to be divided by $\rho$ (the fluid density) to retrieve $u$ on its own.
Calculation of a square-root is needed for calculating the viscosity using Sutherland's law \cite{white_ViscousFluidFlow_1991}, which reads
\begin{equation}
  \mu(T) = T^{\frac{3}{2}} \frac{1 + S}{T + S}
\end{equation}
where $\mu$ is the temperature dependent viscosity, $T$ is the temperature, and $S$ is Sutherland's constant.

\section{Results} \label{sec:results}

\subsection{Validation} \label{sec:validation}

The entire QICFD solver is validated against the classical CFD solver HiPSTAR using low resolution Taylor--Green Vortex (TGV) test cases \cite{giangaspero_CaseC33TaylorGreen_}.
The TGV case is a challenging CFD test case as it is a transient turbulent case, where large structures break down into smaller scales over time.
TGV cases require fine grids to resolve the wide range of spatial scales, and accurate time steps to minimise accumulating errors.
The simulation parameters are $\re = 800$, time step $\Delta t = 0.01$, domain size $L = 2\pi$ in all dimensions, $32^3$ grid size, with $\ma = 0.8$ for the high Mach number case and $\ma = 0.1$ for the low Mach number case.
These test cases are very poorly resolved in both time and space for this $\re$ in classical CFD, and are only meant to show the numerical agreement between QICFD and HiPSTAR.
However, this does not detract from the accuracy requirements of the test case when comparing to the solution in HiPSTAR.
The initial condition for the TGV is
\begin{equation}
  \begin{aligned}
    \rho(x,y,z) &= 1 \\
    u(x,y,z) &= \sin(x) \cos(y) \cos(z) \\
    v(x,y,z) &= - \cos(x) \sin(y) \cos(z) \\
    w(x,y,z) &= 0 \\
    T(x,y,z) &= 1
  \end{aligned}
\end{equation}
\enquote{Tensor cross interpolation} (TCI) \cite{fernandez_LearningTensorNetworks_2025} is used to create the initial TTS, which is a very efficient method for this case, where the initial TTS is only $\chi=3$.
TCI allows a TTS to be efficiently constructed from the function definitions directly, bypassing the need to first create a dense representation, resulting in fast and memory efficient initialisation.
The settings for the QICFD solver specific to the TT format are: relative singular value cutoff $\sigma_c = \num{e-12}$ and $\chi = 256$ for intermediate calculations.

The total kinetic energy and enstrophy are used as metrics to validate the results.
These higher-order metrics are very sensitive to numerical errors, so they can highlight the numerical agreement (and differences) between QICFD and HiPSTAR.
Total kinetic energy is defined as
\begin{equation}
  \mathrm{E_k}(t) = \int_\Omega \frac{1}{2} \rho \, \mathbf{u} \cdot \mathbf{u} \odif{\Omega}
\end{equation}
which is the integrated kinetic energy at each time instance.
Enstrophy is defined as
\begin{equation}
  \zeta (t) = \int_\Omega \frac{1}{2} \rho \, \bm{\omega} \cdot \bm{\omega} \odif{\Omega}
\end{equation}
which is the integrated vorticity magnitude at each time instance.
These quantities can be calculated directly in the TT format using integration \cite{garcia-ripoll_QuantuminspiredAlgorithmsMultivariate_2021}.
Combined with initialisation using TCI, the solver is able to perform the initialisation, simulation, and post-processing entirely within the TT format.
This demonstrates an \enquote{end-to-end} workflow for the QICFD solver, where the solver maintains the TT format at all stages, starting at initialisation and ending at post-processing and reporting.

\begin{figure}
    \centering
    \adjincludegraphics{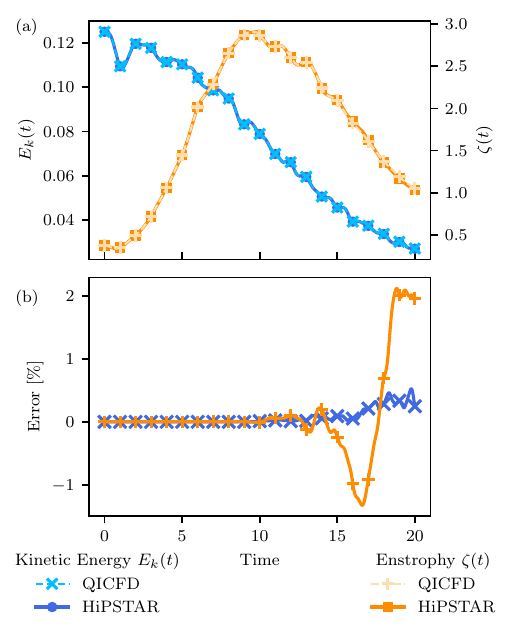}

    \caption{
        Taylor-Green vortex validation test case results, showing (a) the total kinetic energy and enstrophy and (b) percentage error at simulation time $t \in [0,20]$ for the $\ma=0.8$ case produced by the QICFD solver and HiPSTAR classical reference.
    }
    \label{fig:validation}
\end{figure}

Figure \ref{fig:validation} shows the comparison of the $\ma=0.8$ TGV simulation using QICFD and HiPSTAR.
The $\ma=0.1$ simulation is also performed but not shown, as it is nearly incompressible and not the focus of this work.
In general, QICFD shows excellent agreement with the HiPSTAR reference for most of the simulation for both cases, showing below $0.1\%$ error for the total kinetic energy and $0.6\%$ error for enstrophy before $t=15$.
The solution starts to diverge slightly towards the end of the simulations, with the $\ma=0.1$ simulation showing a larger difference to the classical solver.
The error grows to about $2.1\%$ maximum for the $\ma=0.8$ case and $4.6\%$ for the $\ma=0.1$ case towards the end of the simulation.
It is currently unclear what the source of the error is; it may be a combination of numerical errors introduced by the TT format and the chaotic nature of turbulence, which can amplify numerical errors over the course of a simulation.

The TGV test case is a very challenging test case, as small errors can accumulate quickly.
For instance if division was set to a higher tolerance ($\varepsilon =$ \num{e-7} compared to \num{e-9}), the enstrophy results will differ noticeably.
This is also the reason RK4 time stepping and skew-symmetric splitting of derivatives are required.
These features increase the runtime cost per time step by over 4 times, but the solution would become unstable and crash otherwise, even for the classical solver HiPSTAR.
Note that the problem size is sufficiently small such that no final truncation of singular values is necessary, as $\operatorname{max}(\chi) = 128$.
However, intermediate steps can lead to tensors with higher $\chi$ than theoretically allowed, specifically during TTO-TTS contraction and addition.
Truncating these intermediate steps to $\chi=128$ leads to slightly higher errors, hence $\chi=256$ was used.
The agreement between QICFD and HiPSTAR proves the successful and accurate implementation of all algebraic operations in the TT format and its ability to perform compressible CFD simulations, which marks a leap forward for QICFD and expands its range of applications.

\subsection{Simultaneous Simulations} \label{sec:simul}

In the engineering design context, many realisations of similar cases are typically required, for example for testing a design under different operating conditions, or testing many different designs to search for optimal solutions, so a scalable method to perform many parallel simulations can be greatly beneficial, and this can be achieved using QICFD.
By adding an additional coordinate to the TTSs, the QICFD solver can encode additional separate fields in parallel, thus presenting another avenue for better scaling than classical solvers.
\begin{equation}
\begin{array}{c}
\begin{tikzpicture}
  \Vertex[x=0,   y=0,label=$A^{(x,1)}$,position=above,size=0.25,color=blue!30]{A11}
  \Vertex[x=1, y=0,label=$A^{(x,P)}$,position=above,size=0.25,color=blue!30]{A1N}
  \Vertex[x=2,y=0,label=$A^{(y,1)}$,position=above,size=0.25,color=orange!60]{A21}
  \Vertex[x=3,y=0,label=$A^{(y,P)}$,position=above,size=0.25,color=orange!60]{A2N}
  \Vertex[x=4, y=0,label=$A^{(z,1)}$,position=above,size=0.25,color=green!60]{A31}
  \Vertex[x=5, y=0,label=$A^{(z,P)}$,position=above,size=0.25,color=green!60]{A3N}
  \Vertex[x=6, y=0,label=$A^{(c,1)}$,position=above,size=0.25,color=black!60]{A41}
  \Vertex[x=7, y=0,label=$A^{(c,C)}$,position=above,size=0.25,color=black!60]{A4N}

  \draw[line width=1] (A11) -- ++(0,-0.35) node[below] {\scriptsize$q_{x,1}$};
  \draw[line width=1] (A1N) -- ++(0,-0.35) node[below] {\scriptsize$q_{x,P}$};
  \draw[line width=1] (A21) -- ++(0,-0.35) node[below] {\scriptsize$q_{y,1}$};
  \draw[line width=1] (A2N) -- ++(0,-0.35) node[below] {\scriptsize$q_{y,P}$};
  \draw[line width=1] (A31) -- ++(0,-0.35) node[below] {\scriptsize$q_{z,1}$};
  \draw[line width=1] (A3N) -- ++(0,-0.35) node[below] {\scriptsize$q_{z,P}$};
  \draw[line width=1] (A41) -- ++(0,-0.35) node[below] {\scriptsize$q_{c,1}$};
  \draw[line width=1] (A4N) -- ++(0,-0.35) node[below] {\scriptsize$q_{c,C}$};

  \Edge[style={dashed},label={\makebox[1em][c]{$\cdots$}}](A11)(A1N)
  \Edge[style={dashed},label={\makebox[1em][c]{$\cdots$}}](A21)(A2N)
  \Edge[style={dashed},label={\makebox[1em][c]{$\cdots$}}](A31)(A3N)
  \Edge[style={dashed},label={\makebox[1em][c]{$\cdots$}}](A41)(A4N)

  \Edge[lw=1,position=below](A1N)(A21)
  \Edge[lw=1,position=below](A2N)(A31)
  \Edge[lw=1,position=below](A3N)(A41)
\end{tikzpicture}
\end{array},
\label{eqn:tt_diagram_case}
\end{equation}
For example, by adding another tensor a TTS can encode two fields simultaneously, such as two realisations of $u$ (the $x$-velocity field).
This allows the QICFD solver to perform two individual simulations (or more with additional tensors) with the same scaling behaviour as increasing the problem size.
To the authors' knowledge, this work is the first to perform simultaneous simulations using quantum-inspired methods.
To demonstrate this feature, the $\ma=0.8$ TGV test cases are preformed with two and four cases initialised differently.

The alternate initial conditions are
\begin{equation}
    \begin{aligned}
      \rho(x,y,z) &= 1 \\
      u(x,y,z) &= \sin(x) \cos(y) \cos(z) \\
      v(x,y,z) &= - \cos(x) \sin(y) \cos(z) \\
      w(x,y,z) &= w_c \cos(x) \cos(y) \sin(z) \\
      T(x,y,z) &= 1
    \end{aligned}
\end{equation}
which initialised the $w$ velocity field to a non-zero initial condition compared to the previous initial condition.
The two-case simulation uses $w_c \in \{0, 0.1\}$ and the four-case simulation uses $w_c \in \{0, 0.1, 0.2, 0.3\}$.

\begin{figure*}[htb]
    \centering
    \adjincludegraphics{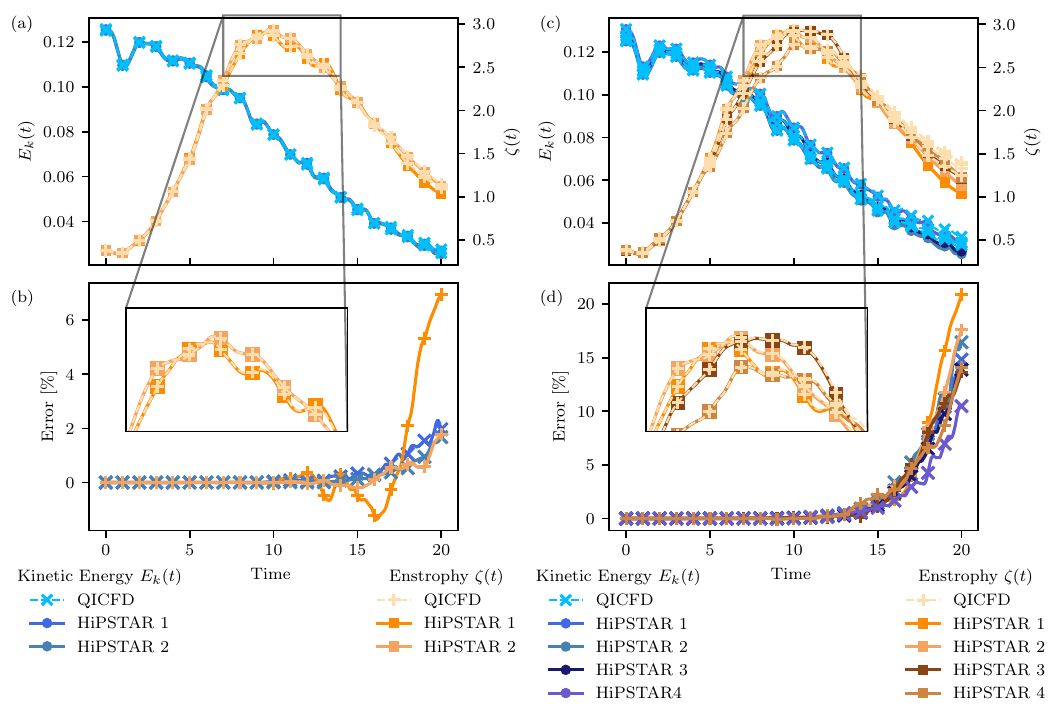}
    \caption{
      Simultaneous simulation validation results for two and four different TGV test cases at $\ma = 0.8$ initialised differently produced by the QICFD solver and HiPSTAR classical reference, showing the total kinetic energy and enstrophy for (a) two cases and (c) four cases; and percentage error for (b) two cases and (d) four cases.
      QICFD results are produced in a single run, while HiPSTAR results are produced in two separate runs.
    } \label{fig:multicase}
\end{figure*}

Figure \ref{fig:multicase} shows the two and four-case simulation results for $\ma=0.8$.
Note the two-case simulation is performed with $\chi = 256$ and the four-case with $\chi=320$ for intermediate calculations, although both have a theoretical maximum of $\chi=256$ for the final TTS.
Again the initial stages of the simulation show very high agreement with the reference classical simulation, and results only start to deviate at later times.
The accuracy of the simultaneous simulation is generally lower than the single case under the same settings, and the four-case simulation is less accurate than the two-case simulation, especially at the later stages.
The runtime of the two-case simulation only increased by $\approx20\%$ (\qty{140}{\second} to \qty{170}{\second}) compared to $200\%$ for a classical solver, while the four-case simulation required \qty{270}{\second}, which is $\approx 200\%$ compared to $400\%$ for a classical simulation.
Note that the four-case simulation used a higher $\chi$ than the two-case simulation, which explains the relatively large increase in runtime, and this is necessary due to the simulation becoming inaccurate and crashing otherwise, the runtime for $\chi=256$ is only \qty{220}{\second} which better matches the expected scaling behaviour if $\chi$ is fixed.
This demonstrates QICFD can perform simultaneous simulations using the same scaling as increasing the problem size, extending the potential advantage of QICFD over classical CFD.
In this work only different initial conditions are demonstrated, however this approach can be easily extended to different boundary conditions and geometries.
This means QICFD could address the needs of engineers and designers for massively parallel simulations, as they may either need to verify their designs in many flow conditions, or perform parallel evaluations of alternate designs.

\subsection{Rank Dependence} \label{sec:rank}

\begin{figure}[htb]
    \centering
    \adjincludegraphics{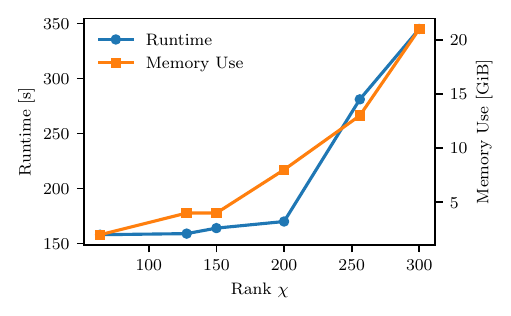}
    \caption{
        Runtime and memory use per step for a $256^3$ grid size $\ma=0.8$ TGV test case with $N=24$ TT tensors saturated with $\chi \in [50, 300]$. Tests are performed on a single GH200 superchip.
    }\label{fig:vrank}
\end{figure}

An increased $\chi$ has large implications on the runtime and memory use of the QICFD solver.
Recall the runtime and memory usage of QICFD scales as $O(N\chi^4)$ and $O(N\chi^3)$ respectively.
Figure \ref{fig:vrank} shows the runtime and memory use of the QICFD solver for different maximum $\chi$ settings for a problem size of $N = 24$, with all TT tensors saturated with this $\chi$.
These tests were performed using a single GH200 superchip.
It is clear that a high $\chi$ is detrimental to the favourable scaling properties of QICFD, as the runtime and memory requirements increase rapidly, suggesting cases that have a high $\chi$ may be unsuitable for QICFD.
One critical question to answer is thus whether complex and industrially-relevant flows have a low $\chi$, and may be thus efficiently represented in the TT format.
To investigate $\chi$ of more complex flows, a-priori analysis is performed on a high-fidelity $\ma=0.8$ TGV case at various $\re$ \cite{pisoni_CompressionSimulationSynthesis_2026, hulst_QuantumInspiredSimulation2D_2026, esmaeili_PrioriAssessmentTensorNetwork_2026}.
Classical simulations using HiPSTAR are first performed, and flow field snapshots generated are encoded into a TTS using SVDs.

\begin{figure}[htb]
    \centering
    \adjincludegraphics{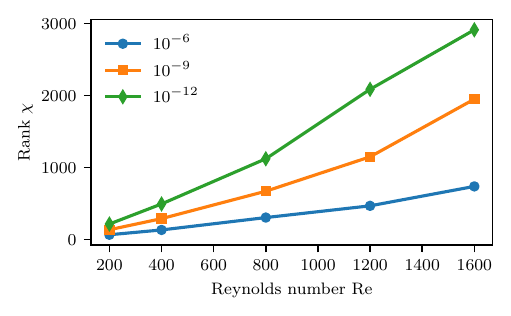}
    \caption{
         Highest $\chi$ for all fields in a high-fidelity $\ma=0.8$ TGV test case for $\re \in [200, 1600]$ throughout all simulation time when encoded as a TTS. Each line represents different $\sigma_c \in \{ \num{e-6}, \num{e-9}, \num{e-12} \}$ during the encoding.
    }
    \label{fig:rank TGV}
\end{figure}

Figure \ref{fig:rank TGV} shows the effects of $\re$ on $\chi$.
The simulation parameters are $\ma = 0.8$, $\Delta t = 0.001, L = 2\pi$ in all dimensions, and $512^3$ grid size.
The Reynolds numbers are $\re \in \{200, 400, 600, 800, 1200, 1600\}$.
For all $\re$ considered, these simulation parameters are considered over-resolved in classical CFD.
Relative singular value cutoffs $\sigma_c \in \{\num{e-6}, \num{e-9}, \num{e-12} \}$ are used during TTS encoding.
For the TGV case, $\chi$ of the problem scales as a power law with $\re$, indicating high $\re$ compressible TGV simulations may prove difficult for QICFD.
In contrast, large computational grid simulations of low $\re$ cases may be very well-suited to QICFD.

$\sigma_c$ significantly impacts the scaling of $\chi$ against $\re$.
It is clear for higher $\sigma_c$, $\chi$ scales slower with $\re$, which can greatly reduce the computational cost of QICFD for any specific problem.
However, higher $\sigma_c$ also leads to lower simulation accuracy, as truncated singular values imply a loss of information, so users can make a trade-off between absolute accuracy and resource requirements through specifying $\sigma_c$.
For many industrial use cases, a certain amount of error can be tolerated, hence a more concrete relationship between $\sigma_c$ and the amount of error introduced would be a useful metric.
However, this may need to be performed on a case-by-case basis, as currently no unified rule exists, and different problems show different relationships between $\sigma_c$, $\chi$ and error.

\begin{figure}[htb]
    \centering
        \adjincludegraphics{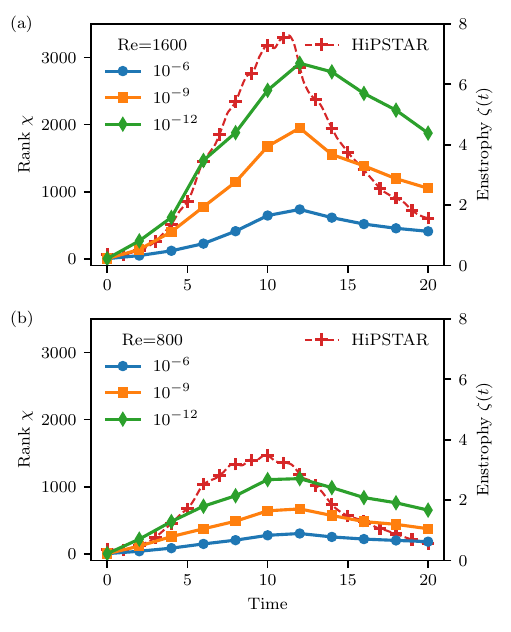}
    \caption{
        The highest $\chi$ of all field TTS at $t \in [0, 20]$ for high-fidelity $\ma=0.8$ TGV simulations at (a) $\re = 1600$ and (b) $\re = 800$ at $\sigma_c \in \{ \num{e-6}, \num{e-9}, \num{e-12} \}$.
        The red plus marked line shows the enstrophy at the corresponding time step to indicate the state of the fluid field.
    }
    \label{fig:rank time}
\end{figure}

Figure \ref{fig:rank time} shows the $\chi$ evolution over time for the $\ma=0.8$ TGV case at $\re \in \{800,1600\}$.
$\chi$ increases over time until around $t=12$ and subsequently decreases.
For QICFD, the peak $\chi$ informs the maximum resource use for a simulation.
The general trend shows that $\chi$ starts extremely low, since the TGV initial condition has a low $\chi$ structure.
$\chi$ subsequently increases as the flow becomes more turbulent and disordered, and generally peaks just over half way through the simulation, where turbulent structures are abundant, and finally subsides as the turbulence decays.
Since the TGV case passes through multiple states of turbulence, in effect the \enquote{local} $\re$ changes over time, and it is clear the different $\re$ has a significant effect on both the actual flow physics and the $\chi$ evolution of the simulation.
The $\re=1600$ case is much more turbulent in general than the $\re=800$ case reflected by the higher enstrophy, and it is apparent that $\chi$ of TTSs needed to encode such fluid fields are correspondingly higher.

These results provide an initial view of how $\chi$ of a TTS is related to the turbulence of the fluid field, however the current results are insufficient to establish a conclusive relationship between $\chi$ and $\re$.
For this high fidelity TGV case at $\re=1600$, the maximum $\chi$ at a $\sigma_c = \num{e-12}$ is very high, up to 2700, which is well out of reach of current QICFD.
For the lowest $\sigma_c = \num{e-6}$, the maximum $\chi$ is 500, which is more manageable but may introduce large errors.
At present the QICFD solver developed in this work is unable to perform this simulation due to poor memory optimisation, hence this setting could not be tested.
With sufficient optimisation and careful memory management, it should be possible to perform simulations with $\chi$ $\approx 500$ using current-generation computing hardware, and this will be a part of future work.

\section{Discussion and Conclusion} \label{sec:discussion}

This work introduces division and square-root algorithms for the TT format, which completes all algebraic operations necessary for compressible CFD simulations in the TT format.
A QICFD solver is developed using the novel algorithms, and is validated by comparison against an in-house state-of-the-art classical CFD solver, HiPSTAR.
Low resolution Taylor--Green Vortex (TGV) test cases at $\re = 800$, $\ma = 0.8$ and $\ma = 0.1$ are performed and compared with HiPSTAR using the total kinetic energy and enstrophy.
QICFD shows excellent agreement with the classical reference, achieving $<0.6\%$ error for the majority of simulation time, confirming the correctness of the new division and square-root algorithms, as well as the QICFD solver in its entirety.
QICFD can also perform simultaneous simulations: TGV simulations are initialised differently and concurrently run with a runtime increase of only $\approx20\%$ for two simultaneous cases and $\approx 200\%$ for four cases (with an increased $\chi$).

However, the scaling behaviour of QICFD depends on $\chi$ of the problem being solved.
A-priori analysis is performed for a high-fidelity TGV test case, and it shows an increasing $\chi$ with $\re$, which suggests this test case is challenging for QICFD. 
Further work is needed to check $\chi$ for other problems to inform QICFD's general applicability.
\cite{hulst_QuantumInspiredSimulation2D_2026} showed that QICFD dynamical simulations with steady statistics at a low $\chi$ settings are still possible, where even if instantaneous snapshots are inaccurate, the overall statistics of the simulation are still correct.
This suggests realistic \enquote{steady state} flows such as channel flows may exhibit different $\chi$ evolutions, as TGV is a transient turbulent case.
If low $\chi$ simulations can still produce satisfactory results even if they would be considered inaccurate when comparing with classical simulations at individual time steps, these steady state simulations may be a method to extend the types of problems that are efficient for QICFD.

Another approach that may prove advantageous could be RANS or LES for QICFD, as these methods reduce the range of scales present in the problem, and should have a lower $\chi$ requirement if the current observations hold.
However their implementation would introduce additional algorithmic challenges due to at-present missing operations in the TT format (depending on the turbulence model used).
It is also clear that further optimisation of the solver, as well as additional features for the QICFD solver are necessary to achieve practical relevance.
Performance optimisation will allow the QICFD solver to perform high $\chi$ simulations for higher accuracy, or accelerate low $\chi$ simulations to reduce the computational overhead of the method, thus reducing the crossover point compared to classical CFD.
Additional solver features such as wall boundary conditions and non-uniform geometry will be necessary for more complex test cases beyond a triply periodic box.
Methods for achieving these features are already present in the incompressible QICFD context \cite{kiffner_TensorNetworkReduced_2023, kornev_TetraFEMNumericalSolution_2024, hulst_QuantumInspiredTensorNetworkFractionalStep_2025, hulst_QuantumInspiredSimulation2D_2026}, and there are no technical hurdles to integrate them with the current solver.

This works shows QICFD is capable of compressible fluid simulations enabled by novel division and square-root algorithms in the TT format.
This opens the door to additional applications that require the simulation of compressible flows, such as aerospace and acoustic applications, as well as any other tensor network algorithm applications that require division and square-roots.
QICFD has the potential to meet the engineering needs for high-fidelity and massively parallel CFD simulations using its favourable scaling over classical CFD for both problem size and parallel simulations.
Therefore QICFD may present a new method for next-generation CFD which bypasses the classical limitations in computational cost, and provides a pathway to out-scale the problem requirements of current classical CFD.
To realise this potential, the $\chi$ characteristics of practical use cases needs to be investigated.
Additionally, incorporation of solver features such as boundary conditions and complex geometries is a natural next step towards industrial relevance.
\\ \\
\noindent
\textbf{Note added at the pre-print submission: }During the final preparation of this manuscript, a pre-print describing similar algorithms for TT division and square-roots was released \cite{wang_IterativeTensorNetwork_2026}.
We strongly encourage the interested reader to consult both works for completeness.

\section*{Acknowledgements}
S.X.M.L. was supported by the Commonwealth through an Australian Government Research Training Program Scholarship [DOI: \href{https://doi.org/10.82133/C42F-K220}{https://doi.org/10.82133/C42F-K220}] and by The University of Melbourne's Elizabeth and Vernon Puzey Scholarship.
This research was supported by The University of Melbourne’s Research Computing Services and the Petascale Campus and by resources provided by the Pawsey Supercomputing Research Centre’s Setonix Supercomputer (\href{https://doi.org/10.48569/18sb-8s43}{https://doi.org/10.48569/18sb-8s43}), with funding from the Australian Government and the Government of Western Australia, via the National Computational Merit Allocation Scheme.

\bibliography{bibtex/cfd,bibtex/qi,bibtex/quantum_papers}

\clearpage
\newpage
\widetext
\appendix

\section*{Appendices for \enquote{Quantum-Inspired Computational Fluid Dynamics for Transient Turbulent Compressible Flows}}

\input{./appendices/symbols.tex}
\input{./appendices/nseq.tex}

\input{./appendices/fdtto.tex}

\end{document}

%% file: figures/summary/summary.tex
\begin{figure*}[htb]
    \centering
    \begin{tikzpicture}[remember picture]
        \node at (1.5, 4.5) {
            \begin{tikzpicture}
                \draw [rounded corners, thick] (-1.5, 0.4) rectangle ++(3, 9);
                \node [align=center] at (0, 8.9) {\textbf{Initial} \\ \textbf{conditions}};
                \node [anchor=west] at (-1.5, 0.6) {\scriptsize $w$-velocity};
                \node (ic3) at (0, 0.9) {case n};
                \draw [dotted, very thick] (0, 1.2) -- (0, 1.5);
                \node (ic2) at (0, 2.9) {\includegraphics[width=2cm]{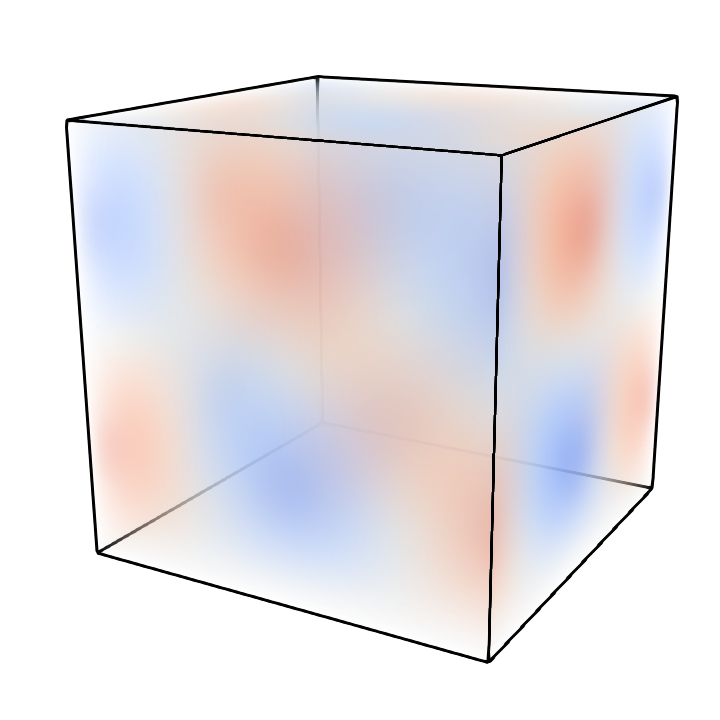}};
                \node at (0, 1.8) {case 3};
                \node (ic1) at (0, 5.2) {\includegraphics[width=2cm]{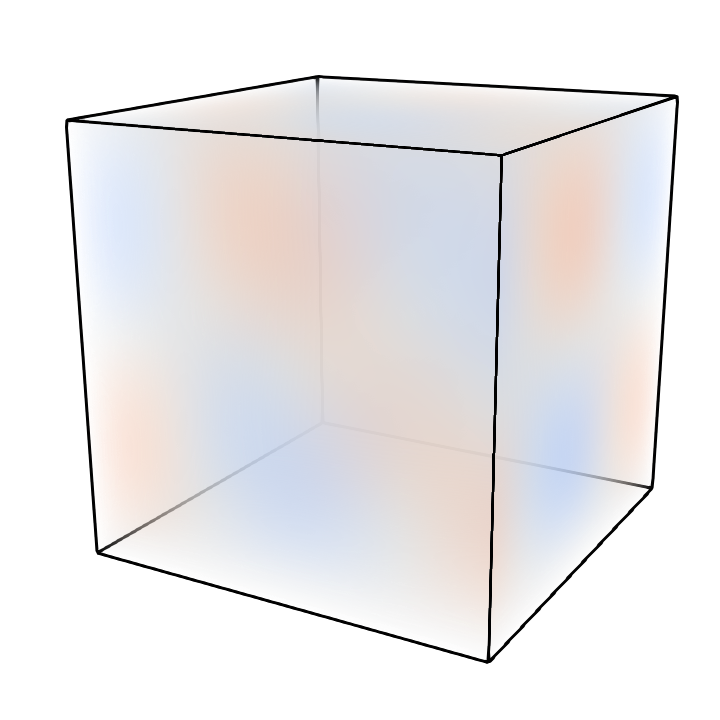}};
                \node at (0, 4.1) {case 2};
                \node (ic0) at (0, 7.5) {\includegraphics[width=2cm]{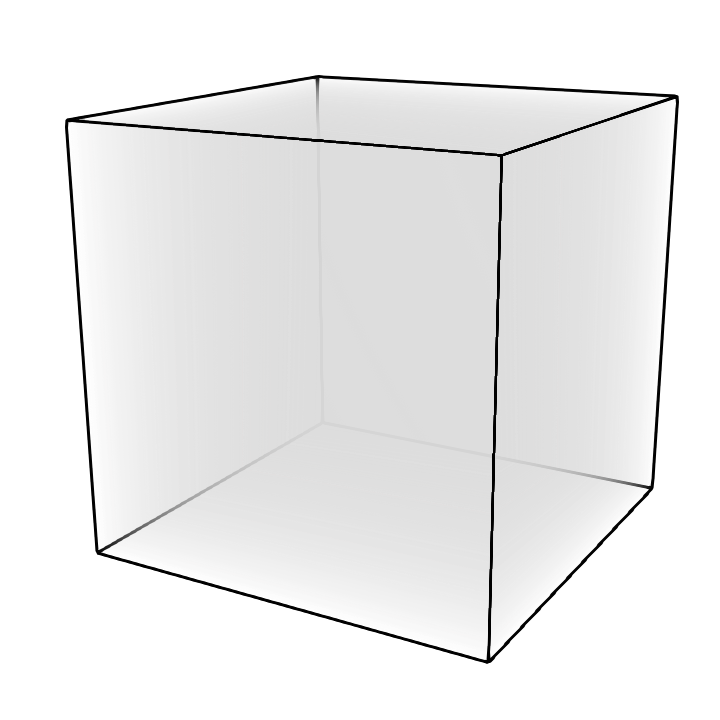}};
                \node at (0, 6.4) {case 1};
            \end{tikzpicture}
        };

        \node (TE) at (9, 2.25){
            \begin{tikzpicture}
                \draw [->, very thick] (-4.5, 1.8) -- node [above] {\textbf{Simultaneous Transient Simulations}} ++(9, 0);
                \draw [rounded corners, thick] (-5, -2) rectangle ++(10, 4.5);
                \node [anchor=west] at (-5, -1.8) {\scriptsize Density \hphantom{2em}};

                \node (t0)  at (-3, -1.7) {$t = 0$ \hphantom{2em}};
                \node (t10) at ( 0, -1.7) {$t = 10$ \hphantom{2em}};
                \node (t20) at ( 3, -1.7) {$t = 20$ \hphantom{2em}};

                \node (w3) at (0, 0.6) {};
                \draw [dotted, very thick]($(t0  |- w3) - (0.8, -0.4)$) -- ++(-0.3, 0.3);
                \draw [dotted, very thick]($(t10 |- w3) - (0.8, -0.4)$) -- ++(-0.3, 0.3);
                \draw [dotted, very thick]($(t20 |- w3) - (0.8, -0.4)$) -- ++(-0.3, 0.3);

                \node (w2) at (0, 0.2) {};
                \node (w2t0)  at ($(t0  |- w2) - (0.3, 0)$) {\includegraphics[width=2cm]{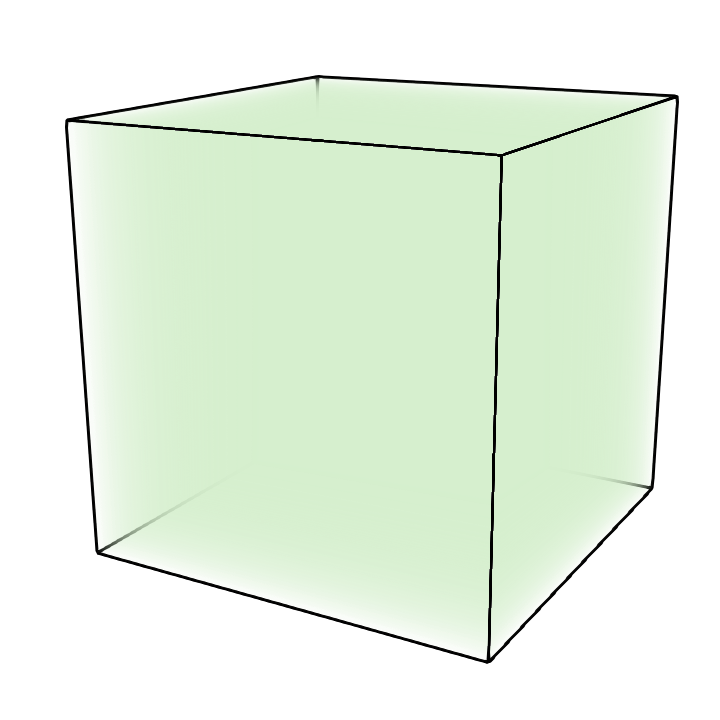}};
                \node (w2t10) at ($(t10 |- w2) - (0.3, 0)$) {\includegraphics[width=2cm]{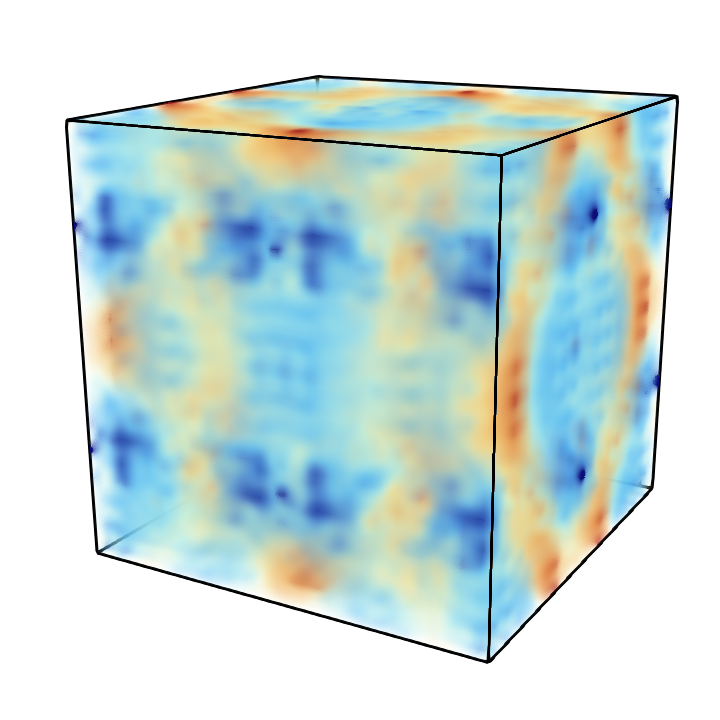}};
                \node (w2t20) at ($(t20 |- w2) - (0.3, 0)$) {\includegraphics[width=2cm]{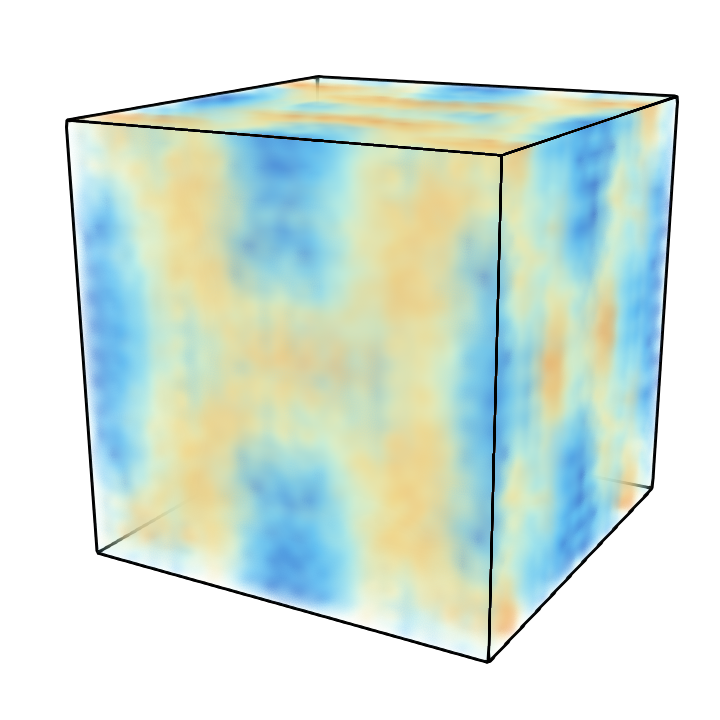}};

                \node (w1) at (0, -0.2) {};
                \node (w1t0)  at ($(t0  |- w1) + (0.0, 0)$) {\includegraphics[width=2cm]{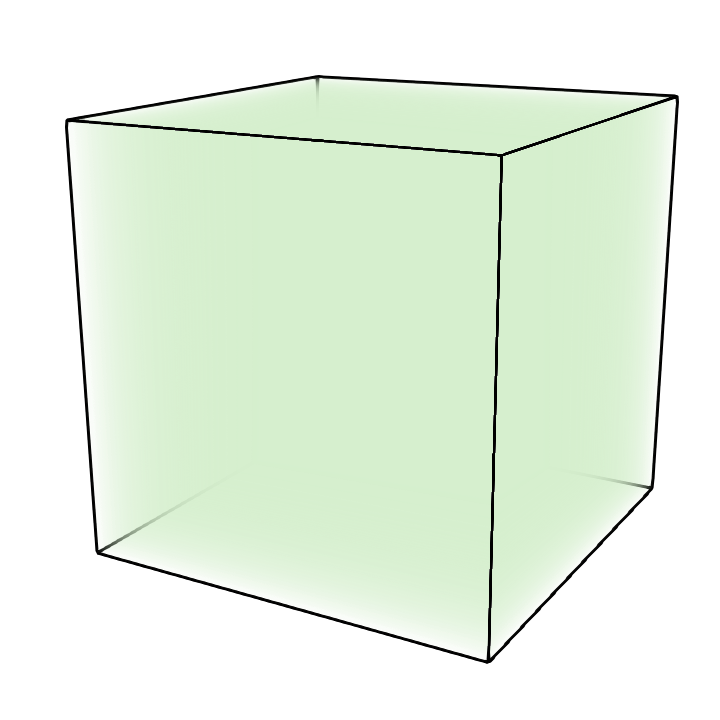}};
                \node (w1t10) at ($(t10 |- w1) + (0.0, 0)$) {\includegraphics[width=2cm]{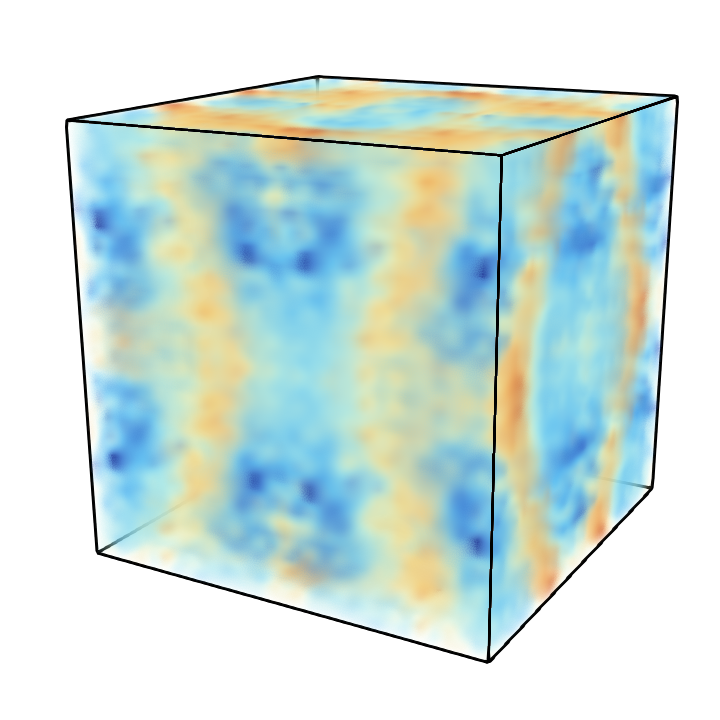}};
                \node (w1t20) at ($(t20 |- w1) + (0.0, 0)$) {\includegraphics[width=2cm]{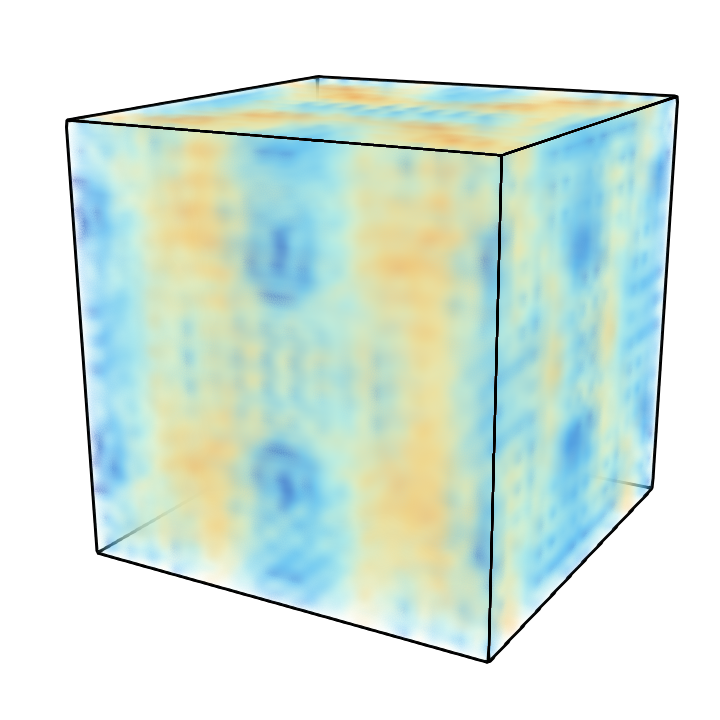}};

                \node (w0) at (0, -0.6) {};
                \node (w0t0)  at ($(t0  |- w0) + (0.3, 0)$) {\includegraphics[width=2cm]{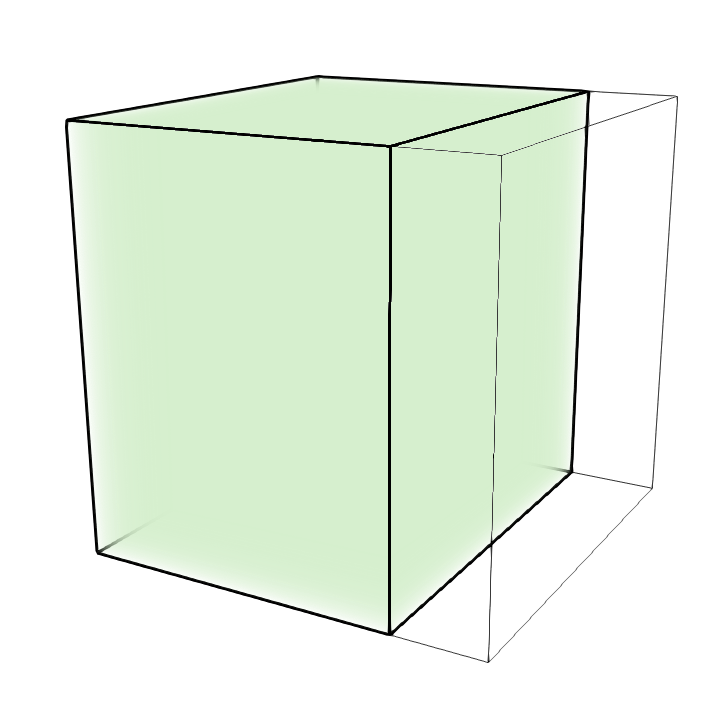}};
                \node (w0t10) at ($(t10 |- w0) + (0.3, 0)$) {\includegraphics[width=2cm]{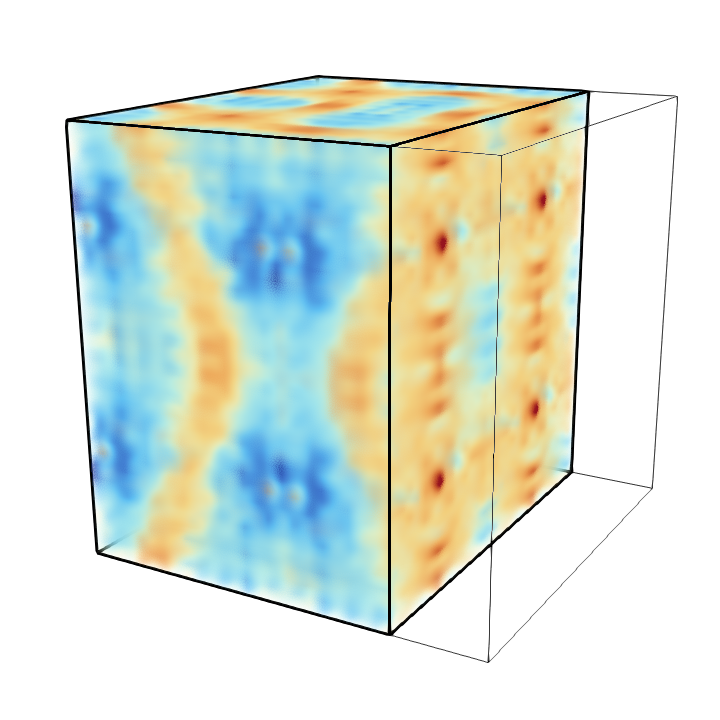}};
                \node (w0t20) at ($(t20 |- w0) + (0.3, 0)$) {\includegraphics[width=2cm]{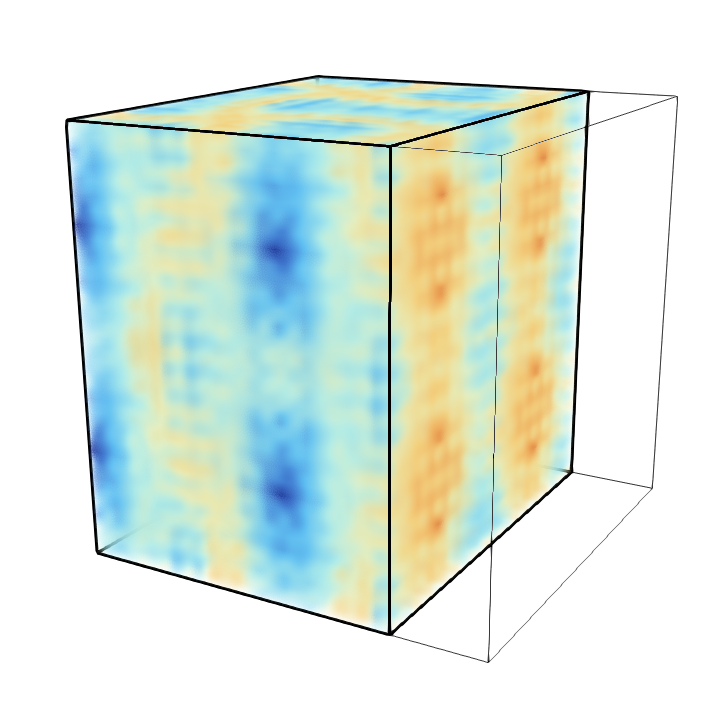}};

            \end{tikzpicture}
        };

        \node (TTS) at (9, 7) {
            \begin{tikzpicture}
                \draw [rounded corners, thick] (-1.25, -1.4) rectangle ++(10, 2.5);
                \node at (3.75, 0.7) {\textbf{Tensor Train Form}};

                \Vertex[x=0,   y=0,size=0.25,color=blue!30]{A11}
                \Vertex[x=1.2, y=0,size=0.25,color=blue!30]{A1P}
                \Vertex[x=2.1,y=0,size=0.25,color=orange!60]{A21}
                \Vertex[x=3.3,y=0,size=0.25,color=orange!60]{A2P}
                \Vertex[x=4.2, y=0,size=0.25,color=green!60]{A31}
                \Vertex[x=5.4, y=0,size=0.25,color=green!60]{A3P}
                \Vertex[x=6.3, y=0,size=0.25,color=black!60]{A41}
                \Vertex[x=7.5, y=0,size=0.25,color=black!60]{A4C}

                \draw[line width=1] (A11) -- ++(0,-0.35) node[below] {};
                \draw[line width=1] (A1P) -- ++(0,-0.35) node[below] {};
                \draw[line width=1] (A21) -- ++(0,-0.35) node[below] {};
                \draw[line width=1] (A2P) -- ++(0,-0.35) node[below] {};
                \draw[line width=1] (A31) -- ++(0,-0.35) node[below] {};
                \draw[line width=1] (A3P) -- ++(0,-0.35) node[below] {};
                \draw[line width=1] (A41) -- ++(0,-0.35) node[below] {};
                \draw[line width=1] (A4C) -- ++(0,-0.35) node[below] {};

                \Edge[style={dashed},label={\makebox[1em][c]{$\cdots$}}](A11)(A1P)
                \Edge[style={dashed},label={\makebox[1em][c]{$\cdots$}}](A21)(A2P)
                \Edge[style={dashed},label={\makebox[1em][c]{$\cdots$}}](A31)(A3P)
                \Edge[style={dashed},label={\makebox[1em][c]{$\cdots$}}](A41)(A4C)

                \Edge[lw=1,position=below](A1P)(A21)
                \Edge[lw=1,position=below](A2P)(A31)
                \Edge[lw=1,position=below](A3P)(A41)

                \draw [decorate, decoration={brace,mirror,raise=14pt}, very thick] ($(A11) + (-2pt,0)$) -- node [below, yshift=-16pt] {$x$} ($(A1P) + (2pt,0)$);
                \draw [decorate, decoration={brace,mirror,raise=14pt}, very thick] ($(A21) + (-2pt,0)$) -- node [below, yshift=-16pt] {$y$} ($(A2P) + (2pt,0)$);
                \draw [decorate, decoration={brace,mirror,raise=14pt}, very thick] ($(A31) + (-2pt,0)$) -- node [below, yshift=-16pt] {$z$} ($(A3P) + (2pt,0)$);
                \draw [decorate, decoration={brace,mirror,raise=14pt}, very thick] ($(A41) + (-2pt,0)$) -- node (case) [below, yshift=-16pt] {case} ($(A4C) + (2pt,0)$);
            \end{tikzpicture}
        };

        \node (TT arith) at (16,7) {
            \begin{tikzpicture}
                \node at (0,0) {
                    \begin{tabular}{c}
                        \textbf{TT Arithmetic} \\
                        Addition \\
                        Multiplication \\
                        Division \\
                        Square-root
                    \end{tabular}
                };
                \draw [green, thick] (1,-0.18) rectangle ++(-2,-0.75);
                \draw [thick, rounded corners] (-1.5,-1.25) rectangle ++(3, 2.5);
            \end{tikzpicture}
        };

        \node at (16, 2.22) {
            \begin{tikzpicture}
                \node at (0.75, 0.5) {\textbf{Results extraction}};

                \Vertex[x=0, y=0, size=0.1,color=blue!30]{a11}
                \Vertex[x=0.5, y=0, size=0.1,color=orange!60]{a21}
                \Vertex[x=1, y=0, size=0.1,color=green!60]{a31}
                \Vertex[x=1.5, y=0, size=0.1,color=black!60]{a41}
                \Vertex[x=0,   y=-0.5, size=0.1,color=white!60]{b11}
                \Vertex[x=0.5, y=-0.5, size=0.1,color=white!60]{b21}
                \Vertex[x=1,   y=-0.5, size=0.1,color=white!60]{b31}
                \Vertex[x=1.5, y=-0.5, size=0.1,color=white!60]{b41}

                \Edge (a11)(a21)
                \Edge (a21)(a31)
                \Edge (a31)(a41)
                \Edge (b11)(b21)
                \Edge (b21)(b31)
                \Edge (b31)(b41)

                \Edge (a11)(b11)
                \Edge (a21)(b21)
                \Edge (a31)(b31)
                \Edge (a31)(b31)
                \Edge (a41)(b41)

                \node at (0.75, -2.5) {
                    \includegraphics{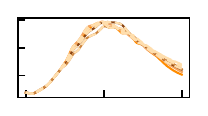}
                };

                \node at (0.75, -1.2) {
                    \begin{NiceTabular}{c}
                        $E_k(t) =\int \rho \mathbf{u} \cdot \mathbf{u} \odif{\Omega}$ \\
                        $\,\,\,\zeta_k(t) = \int \rho \bm{\omega} \cdot \bm{\omega} \odif{\Omega}$
                    \end{NiceTabular}
                };

                \draw [rounded corners, thick] (-1, -3.5) rectangle ++(3.5, 4.5);
            \end{tikzpicture}
        };


        \node [minimum size=0] (ic vert) at ($(3.5, 0 |- ic0) + (0, 1.7)$) {};
        \draw [->, thick] (ic vert.center) -- (ic vert.center -| TTS.north) node [midway, below] {Tensor Cross Interpolation} -- (TTS.north);
        \draw [thick] (ic0 -| 3.1,0) -| (ic vert.center);
        \draw [thick] (ic1 -| 3.1,0) -- (ic1 -| ic vert);
        \draw [thick] (ic2 -| 3.1,0) -- (ic2 -| ic vert);
        \draw [thick] (ic3 -| 3.1,0) -| (ic vert);

        \draw [->, thick] (TTS.south) -- node [midway, left=1em, align=center, draw=green] (eq arrow) {\textbf{Compressible} \\ \textbf{Navier--Stokes equations}} (TE.north);
        \draw [->, thick] (TT arith) |- ($(eq arrow -| TTS.south) + (0.1,0)$);
    \end{tikzpicture}
    \caption{
        Overview figure:
        The overall workflow for an end-to-end Quantum Inspired Computational Fluid Dynamics (QICFD) solver for compressible flows.
        Multiple initial conditions ($w$-velocity shown, colours indicate the magnitude and sign) are encoded into the Tensor Train (TT) form using Tensor Cross Interpolation, where each TT is capable of encoding all initial conditions simultaneously.
        The QICFD solver then performs CFD calculations using the compressible Navier--Stokes equations using TT arithmetic algorithms, and all instances of the different initial conditions can be simulated simultaneously at marginal additional cost over a single instance (density fields at time $t \in \{0,10,20\}$ shown, colours indicate the magnitude).
        High-order statistics, such as total kinetic energy $E_k(t)$ and enstrophy $\zeta(t)$ can be extracted prom the simulation results within the TT form as a post processing step.
        Our QICFD solver is the first complete compressible Navier-Stokes equation solver, enabled by novel TT division and square-root algorithms developed in this work, and it never requires constructing the full dense representation of the data, which allows it to benefit from the advantageous scaling of the TT form from start to finish.
    }
    \label{fig:summary}
\end{figure*}

%% file: appendices/symbols.tex
\section{Symbols and nomenclature}

 \begin{table*}[htb]
    \centering
    \begin{NiceTabular}{l|p{5cm}|p{10cm}}
        Symbol & Full name & Additional comments \\
        \hline \hline
         & Dense form & Original classical form of data \\
        TT & Tensor Train & Refers to both TTS and TTO \\
        TTS & Tensor Train State & TT form of a vector (data) (Matrix Product State (MPS) in MPS context) \\
        TTO & Tensor Train Operator & TT form of a matrix (linear operator) (Matrix Product Operator (MPO) in MPS context) \\
        $D$ & Number of spatial dimensions & \\
        $P$ & Number of binary indices per dimension & \\
        $A^n$ & TT tensor & Tensor $n$ in TT \\
        $N$ & Total number of TT tensors & $N = PD$ \\
        $n$ & & Index of tensor in TT, $n \in [1, N]$ \\
        $q_{n}$ & Physical index & Physical index $n$ (or dimension of) (site in MPS context) \\
        $\alpha_n$ & Virtual index & Virtual index $n$ (or dimension of) (bond in MPS context) \\
        $\sigma$ & Singular values & Singular values generated by Singular Value Decomposition (SVD) \\
        $\sigma_c$ & Relative singular value cutoff & Cutoff relative to the largest singular value during SVD truncation \\
        $\chi$ & Rank & Maximum dimension of $\alpha_n$ (bond dimension in MPS context) \\
        $x$ & $x-$spatial dimension & \\
        $y$ & $y-$spatial dimension & \\
        $z$ & $z-$spatial dimension & \\
        $\re$ & Reynolds number & Ratio of inertial forces over viscous forces \\
        $\ma$ & Mach number & Characteristic flow velocity over speed of sound \\
        $\odot$ & Element-wise multiply & \\
        $\mathbf{D}$ & Input TTS & Divisor in division, input state in square-root \\
        $\mathbf{N}$ & Numerator TTS & Used in division \\
        $a$ & Lower bound of $\mathbf{D}$ & $a \leq \operatorname{minimum}(\mathbf{D})$\\
        $b$ & Upper bound of $\mathbf{D}$ & $b \geq \operatorname{maximum}(\mathbf{D})$\\
        $\varepsilon$ & Error metric & Used in division and square-root \\
        $i$ & Iteration counter & Used in division and square-root \\
        $t$ & Time & \\
        $\rho$ & Density & \\
        $u$ & $x-$velocity & \\
        $v$ & $y-$velocity & \\
        $w$ & $z-$velocity & \\
        $e$ & Specific energy & \\
        $T$ & Temperature & \\
        $p$ & Pressure & \\
        $\mu$ & Viscosity & \\
        $\tau$ & Viscous stresses & \\
        $q$ & Heat fluxes & \\
        $\Pr$ & Prandtl number & \\
        $S$ & Sutherland's constant & \\
        $\gamma$ & Adiabatic constant & \\
        $\alpha_q, \beta_q, \alpha, \gamma$ & Skew-symmetric coefficients & Note $\gamma$ is also used for the adiabatic constant \\
        $\nabla$ & Differential operator & \\
        $E_k(t)$ & Total kinetic energy & \\
        $\zeta(t)$ & Enstrophy & \\
        $\Delta t$ & Time step & \\
        $L$ & Domain size in one dimension & \\
    \end{NiceTabular}
    \caption{Full table of symbols and nomenclature}
    \label{tab:full symbols}
\end{table*}

%% file: appendices/nseq.tex
\section{QICFD Solver Details} \label{sec:solver details}

This section will detail the QICFD solver.
The main elements of the QICFD solver are:
\begin{itemize}
  \item Direct numerical simulation
  \item Standard RK4 time stepping
  \item Forth order finite difference for spatial derivatives
  \item Sutherland's law for viscosity \cite{white_ViscousFluidFlow_1991}
  \item Skew-symmetric splitting of derivatives \cite{kennedy_ReducedAliasingFormulations_2008}
  \item Uniform Cartesian grid
  \item Periodic boundary conditions
\end{itemize}

This solver performs direct numerical simulations (DNS), which mean the full Navier--Stokes equations are used without any modelling, and can be considered the most accurate type of CFD simulation.
The Navier--Stokes equations for a compressible flow in conservative derivative form with skew-symmetric splitting of derivatives \cite{kennedy_ReducedAliasingFormulations_2008}, are
\begin{gather}
    \pdv{\rho}{t} = - \nabla_j \cdot \left( \alpha_q \rho \mathbf{u}_j \right)
    - \beta_q \left( \rho \nabla_j \cdot \mathbf{u}_j + \mathbf{u}_j \cdot \nabla_j \rho \right) \\
    \pdv{\rho \mathbf{u}_i}{t} = - \nabla_j \cdot \left( \alpha \rho \mathbf{u}_i \mathbf{u}_j + p \delta_{ij} - \mathbf{\tau}_{ji} \right)
    - \gamma \left( \mathbf{u}_i \mathbf{u}_j \cdot \nabla_j \rho + \rho \mathbf{u}_j \cdot \nabla_j \mathbf{u}_i + \rho \mathbf{u}_i \nabla_j \cdot \mathbf{u}_j \right) \\
    \pdv{\rho e}{t} = - \nabla_j \cdot \left( \alpha \rho e \mathbf{u}_j + \alpha_q p \mathbf{u}_{j} - \mathbf{\tau}_{ji} \cdot \mathbf{u}_{i} + \mathbf{q}_j \right)
    - \gamma \left( e \mathbf{u}_j \cdot \nabla_j \rho + \rho \mathbf{u}_j \cdot \nabla_j e + \rho e \nabla_j \cdot \mathbf{u}_j \right)
    - \beta_q \left( p \nabla_j \cdot \mathbf{u}_j + \mathbf{u}_j \cdot \nabla_j p \right)
\end{gather}
$\rho$ is the fluid density, $\mathbf{u}$ is the fluid velocity vector, and $e$ the fluid specific energy, $\mathbf{\tau}$ is the viscous stress tensor, $\mathbf{q}$ is the heat flux vector, and $p$ is the fluid pressure, $\nabla$ is the derivative operator, $\delta$ is the Kronecker delta, $i,j$ are spatial direction indices, $\alpha_q, \beta_q, \alpha, \gamma$ are coefficients, and terms with coefficient of $\beta$ are omitted as it is set to $\beta=0$ in this work \cite{kennedy_ReducedAliasingFormulations_2008}.

The Runge-Kutta 4 (RK4) time stepping scheme is an forth-order accurate explicit time stepping scheme that calculates the time derivative.
For a differential equation that reads
\begin{equation}
    \odv{y}{t} = f(t, y)
\end{equation}
with time step $\Delta t$, the pseudo time derivatives at step $n$ are calculated using
\begin{align}
    k_1 &= f(t_n, y_n) \\
    k_2 &= f \left( t_n + \frac{\Delta t}{2}, y_n + k_1 \frac{\Delta t}{2} \right) \\
    k_3 &= f \left( t_n + \frac{\Delta t}{2}, y_n + k_2 \frac{\Delta t}{2} \right) \\
    k_4 &= f \left( t_n + \Delta t, y_n + k_3 \Delta \right) \\
\end{align}
then $y_{n+1}$ is updated using
\begin{equation}
    y_{n+1} = y_n + \frac{\Delta t}{6}\left( k_1 + 2 k_2 + 2 k_3 + k_4 \right)
\end{equation}
The actual equations for $f(t, y)$ are described above and in Section \ref{sec:ns eq}.

Finite difference solvers discretise the problem on a structured grid, in this work a uniform 3D Cartesian grid, and quantities are calculated at the nodes of the grid.
Spatial derivatives are discretised using fourth-order central difference stencils, which read as:
\begin{equation}
    \left. \pdv{u}{x} \right|_{x = i} \approx \frac{-u_{i+2} + 8 u_{i+1} - 8u_{i-1} + u_{i-2}}{12 \Delta x}
\end{equation}
where $u$ is the quantity for which the derivative is sought and $i$ is the index of the grid cell.
Forth order is the default for HiPSTAR \cite{sandberg_CompressibleDirectNumerical_2015}, however up to eighth order is supported.
The implementation also automatically implies periodic boundary conditions, see Section \ref{sec:FDTTO}.

\subsection{Compressible Navier--Stokes Equations} \label{sec:ns eq}

The equations solved are presented in a more computationally friendly vectorised form, and they are equivalent to the equations presented by \textcite{kennedy_ReducedAliasingFormulations_2008}, with $\alpha_q = \beta_q = \alpha = \gamma = 0.5$ and $\beta = 0$.
The exact equations solved in vectorised form are
\begin{equation}
    \pdv{\mathbf{Q}}{t} = \pdv{\mathbf{E}}{x} + \pdv{\mathbf{F}}{y} + \pdv{\mathbf{G}}{z} + \mathbf{S}
\end{equation}

$\mathbf{Q}$ are the conservative variables.
\begin{equation}
    \mathbf{Q} = \begin{bNiceMatrix}
        \rho & \rho u & \rho v & \rho w & \rho E
    \end{bNiceMatrix}^T
\end{equation}
$\rho$ is the fluid density, $u,v,w$ are the fluid velocity in the $x,y,z$ directions respectively, and $e$ the fluid specific energy.

$e$ is related to the fluid temperature $T$ and other quantities by
\begin{equation}
    e = \frac{T}{\gamma (\gamma - 1) \mathrm{Ma}^2} + \frac{1}{2} \left( u^2 + v^2 + w^2 \right)
\end{equation}
$\gamma$ is the adiabatic constant and $\mathrm{Ma}$ is the Mach number.

$\mathbf{E},\mathbf{F},\mathbf{G}$ are the flux vectors.
\begin{align}
    \mathbf{E} &= \begin{bNiceMatrix}
        - \alpha_q \rho u \\
        - \alpha \rho u u + \tau_{xx} - p \\
        - \alpha \rho u v + \tau_{xy} \\
        - \alpha \rho u w + \tau_{xz} \\
        - \alpha \rho u e - \alpha_q u p + u \tau_{xx} + v \tau_{xy} + w \tau_{xz} - q_x
    \end{bNiceMatrix} \\
    \mathbf{F} &= \begin{bNiceMatrix}
        - \alpha_q \rho v \\
        - \alpha \rho v u + \tau_{yx} \\
        - \alpha \rho v v + \tau_{yy} - p \\
        - \alpha \rho v w + \tau_{yz} \\
        - \alpha \rho v e - \alpha_q v p + u \tau_{yx} + v \tau_{yy} + w \tau_{yz} - q_y
    \end{bNiceMatrix} \\
    \mathbf{G} &= \begin{bNiceMatrix}
        - \alpha_q \rho w \\
        - \alpha \rho w u + \tau_{zx} \\
        - \alpha \rho w v + \tau_{zy} - p \\
        - \alpha \rho w w + \tau_{zz} \\
        - \alpha \rho w e - \alpha_q w p + u \tau_{zx} + v \tau_{zy} + w \tau_{zz} - q_z
    \end{bNiceMatrix}
\end{align}
$\tau$ are the viscous stresses, $q$ are the heat flux, and $p$ is the fluid pressure.

$\mathbf{S}$ contains the skew symmetric terms
\begin{equation}
    \mathbf{S} = \begin{bNiceMatrix}
        - \beta_q \left( 
            \rho \pdv{u}{x} + \rho \pdv{v}{y} + \rho \pdv{w}{z} +
            u \pdv{\rho}{x} + v \pdv{\rho}{y} + w \pdv{\rho}{z}
        \right) \\
        - \gamma \left(
            \rho u \pdv{u}{x} + \rho v \pdv{u}{y} + \rho w \pdv{u}{z} +
            \rho u \pdv{u}{x} + \rho u \pdv{v}{y} + \rho u \pdv{w}{z} +
            u u \pdv{\rho}{x} + u v \pdv{\rho}{y} + u w \pdv{\rho}{z}
        \right) \\
        - \gamma \left(
            \rho u \pdv{v}{x} + \rho v \pdv{v}{y} + \rho w \pdv{v}{z} +
            \rho v \pdv{u}{x} + \rho v \pdv{v}{y} + \rho v \pdv{w}{z} +
            v u \pdv{\rho}{x} + v v \pdv{\rho}{y} + v w \pdv{\rho}{z}
        \right) \\
        - \gamma \left(
            \rho u \pdv{w}{x} + \rho v \pdv{w}{y} + \rho w \pdv{w}{z} +
            \rho w \pdv{u}{x} + \rho w \pdv{v}{y} + \rho w \pdv{w}{z} +
            w u \pdv{\rho}{x} + w v \pdv{\rho}{y} + w w \pdv{\rho}{z}
        \right) \\
        \begin{aligned}
            - \beta_q &\left(
                p \pdv{u}{x} + p \pdv{v}{y} + p \pdv{w}{z} +
                u \pdv{p}{x} + v \pdv{p}{y} + w \pdv{p}{z}
            \right) + \\
            & \qquad
            \begin{aligned}
                - \gamma \left( \vphantom{\pdv{a}{b}} \right. 
                & u e \pdv{\rho}{x} + v e \pdv{\rho}{y} + w e \pdv{\rho}{z} + \rho e \pdv{u}{x} + \rho e \pdv{v}{y} + \rho e \pdv{w}{z} + \\
                & \frac{1}{\gamma (\gamma - 1) \mathrm{Ma}^2} \rho u \pdv{T}{x} + \rho u u \pdv{u}{x} + \rho u v \pdv{v}{x} +\rho u w \pdv{w}{x} + \\
                &\frac{1}{\gamma (\gamma - 1) \mathrm{Ma}^2}\rho v \pdv{T}{y} + \rho v u \pdv{u}{y} + \rho v v \pdv{v}{y} +\rho v w \pdv{w}{y} + \\
                &\frac{1}{\gamma (\gamma - 1) \mathrm{Ma}^2}\rho w \pdv{T}{z} + \rho w u \pdv{u}{z} + \rho w v \pdv{v}{z} +\rho w w \pdv{w}{z}
                \left. \vphantom{\pdv{a}{b}} \right)
            \end{aligned}
        \end{aligned}
    \end{bNiceMatrix}
\end{equation}

The viscosity $\mu$ is calculated using Sutherland's law \cite{white_ViscousFluidFlow_1991}
\begin{equation}
    \mu(T) = T^{\frac{3}{2}} \frac{1 + S}{T + S}
\end{equation}
$S$ is Sutherland's constant.

The viscous stresses $\tau$ are
\begin{equation}
    \tau_{ik} = \frac{\mu}{\mathrm{Re}} \left( \pdv{u_i}{x_k} + \pdv{u_k}{x_i} - \frac{2}{3} \pdv{u_j}{x_j} \delta_{ik} \right)
\end{equation}

The heat fluxes $q$ are
\begin{equation}
    q_k = \frac{-\mu}{(\gamma - 1) \mathrm{Ma}^2 \mathrm{Pr} \mathrm{Re}} \pdv{T}{x_k}
\end{equation}
$\mathrm{Re}$ is the Reynolds number, $\mathrm{Pr}$ is the Prandtl number.

The equations are closed using the non-dimensional equation of state (ideal gas law)
\begin{equation}
    p = \frac{\rho T}{\gamma \mathrm{Ma}^2}
\end{equation}

%% file: appendices/fdtto.tex
\section{Finite Difference TTOs} \label{sec:FDTTO}

This section will detail the explicit construction of finite difference TTOs.
The construction is an extension of \textcite{gourianov_QuantumInspiredApproach_2022} to include $\pm 4$ neighbouring cells, allowing up to eighth order accuracy.

As a start, a central finite difference operator operating on $\pm 1$ neighbouring cell reads
\begin{equation}
    g(x_i) = \frac{a f(x_{i-1}) + b f(x_i) + c f(x_{i+1})}{\Delta x}
\end{equation}
with coordinate $x$ and coefficients $a,b,c$.

\begin{figure}[ht]
    \centering
    \SetVertexStyle[MinSize=1\DefaultUnit,TextFont=\normalsize]
    \begin{tikzpicture}
        \Vertex[IdAsLabel,Math,size=0.25,color=blue!30,position=above]{L}
        \Vertex[IdAsLabel,Math,x=1.5,size=0.25,color=blue!30,position=above left]{A^{(1)}}
        \Vertex[IdAsLabel,Math,x=4,size=0.25,color=blue!30,position=above left]{A^{(N)}}
        \Vertex[IdAsLabel,Math,x=5.5,size=0.25,color=blue!30, position=above]{R}
        \Vertex[Pseudo,x=1.5,y=1.2]{C1u}
        \Vertex[Pseudo,x=1.5,y=-1.2]{C1d}
        \Vertex[Pseudo,x=4,y=1.2]{C2u}
        \Vertex[Pseudo,x=4,y=-1.2]{C2d}

        \Edge(L)(A^{(1)})
        \Edge[style={dotted}](A^{(1)})(A^{(N)})
        \Edge(A^{(N)})(R)
        
        \Edge(A^{(1)})(C1u)
        \Edge(A^{(1)})(C1d)
        \Edge(A^{(N)})(C2u)
        \Edge(A^{(N)})(C2d)
    \end{tikzpicture}
    \caption{Finite difference TTO. The TT tensors between $A^{(1)}$ and $A^{(N)}$ are omitted for brevity.}
    \label{fig:FDTTO}
\end{figure}
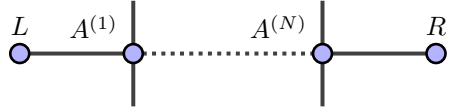

A finite difference TTO will have the form shown in Figure \ref{fig:FDTTO}.
To construct the finite difference TTO for this operation, first consider the case $a=0$, where only the following neighbour is summed.
One can consider each tensor of the TTO as a state machine operating using a binary addition lookup table.
\begin{table}[ht]
    \centering
    \begin{NiceTabular}{|c|c|c|c|}
        \hline
         input 1 & input 2 & output & carry \\ \hline \hline
         0 & 0 & 0 & 0 \\ \hline
         1 & 0 & 1 & 0 \\ \hline
         0 & 1 & 1 & 0 \\ \hline
         1 & 1 & 0 & 1 \\ \hline
    \end{NiceTabular}
    \caption{Binary addition lookup table}
    \label{tab:p1}
\end{table}

Then the tensors $A^{(n)}$ where $1 \leq n \leq N$ are (using 0 based indexing)
\begin{equation}
    \begin{aligned}
        A[0, 0, 0, 0] &= 1 \\
        A[1, 0, 1, 0] &= 1 \\
        A[0, 1, 1, 0] &= 1 \\
        A[1, 1, 0, 1] &= 1
    \end{aligned}
\end{equation}
with all other entries $0$.

The index ordering is [up, left, down, right], corresponding to [input 1, input 2, output, carry].
Up corresponds to the input from a TTS this operator is applied to, left is input 2 (which is the carry in from the previous TT tensor), down is the output of the TTS once the TTO is contracted, and right is carry out which is connected to the next TT tensor.
An additional note is the TTS to contract has binary physical index ordering from left to right, such that $x = 2^0 q_1 + 2^1 q_2 + \cdots + 2^{N-1} q_N$.
\begin{figure}[ht]
    \centering
    \SetVertexStyle[MinSize=1\DefaultUnit,TextFont=\normalsize]
    \begin{tikzpicture}
        \Vertex[label=$A^{(n)}$, color=blue!30]{A}
        \Vertex[Pseudo,label=left,x=-1.5]{L}
        \Vertex[Pseudo,label=up,y=1.5]{U}
        \Vertex[Pseudo,label=down,y=-1.5]{D}
        \Vertex[Pseudo,label=right,x=1.5]{R}

        \Edge(A)(L)
        \Edge(A)(U)
        \Edge(A)(D)
        \Edge(A)(R)
    \end{tikzpicture}
    \caption{Index labelling of finite difference TTO tensor}
    \label{fig:FDTTO notation}
\end{figure}
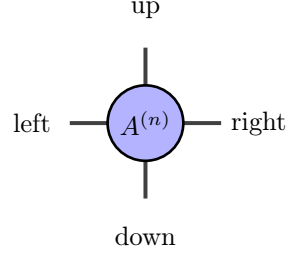

Then the vector $L$ can contain the coefficients $[b, c]$, which are the coefficients associated with a carry of 0 and 1 respectively, and $R$ contains $[1, 1]$, which sums the results of all carry states.

To introduce $a \neq 0$, and additional state corresponding to binary subtraction is added, with total addition table
 \begin{table}[ht]
    \centering
    \begin{NiceTabular}{|c|c|c|c|}
        \hline
         input 1 & input 2 & output & carry \\ \hline \hline
         0 & 0 & 0 & 0 \\ \hline
         1 & 0 & 1 & 0 \\ \hline
         0 & 1 & 1 & 0 \\ \hline
         1 & 1 & 0 & 1 \\ \hline
         0 & -1 & 1 & -1 \\ \hline
         1 & -1 & 0 & 0 \\ \hline
    \end{NiceTabular}
    \caption{Binary addition and subtraction lookup table}
    \label{tab:pm1}
\end{table}

And tensors $A^{(n)}$ $1 \leq n \leq N$ (with index 2 representing carry of -1)
\begin{equation}
    \begin{aligned}
        A[0, 0, 0, 0] &= 1 \\
        A[1, 0, 1, 0] &= 1 \\
        A[0, 1, 1, 0] &= 1 \\
        A[1, 1, 0, 1] &= 1 \\
        A[0, 2, 1, 2] &= 1 \\
        A[1, 2, 0, 0] &= 1
    \end{aligned}
\end{equation}
$L = [b, c, a]$, $R = [1, 1, 1]$

Finally, to accommodate finite differences involving more distant neighbours, additional indices corresponding to $\pm 2$ neighbours or more are added, and also utilise the odd or even symmetry of the finite difference coefficients.
This creates fourth-order accurate central finite difference operators, which is primarily used in this work.
 \begin{table}[ht]
    \centering
    \begin{NiceTabular}{|c|c|c|c|}
        \hline
         input 1 & input 2 & output & carry \\ \hline \hline
         0 & $\pm2$ & 0 & $\pm1$ \\ \hline
         1 & $\pm2$ & 1 & $\pm1$ \\ \hline
    \end{NiceTabular}
    \caption{$\pm 2$ neighbour lookup table, entries for $\pm 0,1$ neighbour omitted}
    \label{tab:pm2}
\end{table}

All TT tensors $2 \leq n \leq N$ are the same as the 1 neighbour case, and only modifying $n=1$ is necessary.
Index 3 now represents an offset of 2, and $s = 1$ or $s = -1$ can be used to incorporate the odd or even symmetry efficiently.
The following entries are added to $A^{(1)}$ (previous entries omitted).
\begin{equation}
    \begin{aligned}
        A^{(1)}[0, 3, 0, 1] &= 1 \\
        A^{(1)}[1, 3, 1, 1] &= 1 \\
        A^{(1)}[0, 3, 0, 2] &= s \\
        A^{(1)}[1, 3, 1, 2] &= s
    \end{aligned}
\end{equation}

Extending to $\pm 4$ neighbours allow up to eighth-order accurate central finite difference, but this is not used in the present work.
With index 4 for $\pm 3$ and 5 for $\pm 4$
 \begin{table}[ht]
    \centering
    \begin{NiceTabular}{|c|c|c|c|}
        \hline
         input 1 & input 2 & output & carry \\ \hline \hline
         0 & 3 & 1 & 1 \\ \hline
         1 & 3 & 0 & 2 \\ \hline
         0 & -3 & 1 & -2 \\ \hline
         1 & -3 & 0 & -1 \\ \hline
         0 & 4 & 0 & 2 \\ \hline
         1 & 4 & 1 & 2 \\ \hline
         0 & -4 & 0 & -2 \\ \hline
         1 & -4 & 1 & -2 \\ \hline
    \end{NiceTabular}
    \caption{$\pm 3,4$ neighbour lookup table, entries for $\pm 0,1,2$ neighbours omitted}
    \label{tab:pm4}
\end{table}

In this case $A^{(2)}$ needs to process $+2$ and $-2$ carry in separately, so the output index is different than the input index.
Input index $1 \rightarrow +1$, $2 \rightarrow -1$, $3 \rightarrow \pm 2$, $4 \rightarrow \pm 3$, $5 \rightarrow \pm 4$.
Output index $1 \rightarrow +1$, $2 \rightarrow -1$, $3 \rightarrow +2$, $4 \rightarrow -2$.
\begin{equation}
    \begin{aligned}
        A^{(1)}[0, 4, 1, 1] &= 1 \\
        A^{(1)}[1, 4, 0, 3] &= 1 \\
        A^{(1)}[0, 4, 1, 4] &= s \\
        A^{(1)}[1, 4, 0, 2] &= s \\
        A^{(1)}[0, 5, 0, 3] &= 1 \\
        A^{(1)}[1, 5, 1, 3] &= 1 \\
        A^{(1)}[0, 5, 0, 4] &= s \\
        A^{(1)}[1, 5, 1, 4] &= s
    \end{aligned}
\end{equation}

Input index $1 \rightarrow +1$, $2 \rightarrow -1$, $3 \rightarrow +2$, $4 \rightarrow -2$.
Output index $1 \rightarrow +1$, $2 \rightarrow -1$.
\begin{equation}
    \begin{aligned}
        A^{(2)}[0, 3, 0, 1] &= 1 \\
        A^{(2)}[1, 3, 1, 1] &= 1 \\
        A^{(2)}[0, 4, 0, 2] &= 1 \\
        A^{(2)}[1, 4, 1, 2] &= 1
    \end{aligned}
\end{equation}

Finally, these finite difference TTO can be constructed for a portion of a TTS to perform spatial derivatives in a single spatial direction.
For example for a 2D TTS with coordinates $q_{x,1} \cdots q_{x,N} q_{y,1} \cdots q_{y,M}$, a TTO of length $N$ contracted with $q_{x,1} \cdots q_{x,N}$ will perform the $x-$spatial derivative.